\documentclass[A4,12pt]{article}

\pdfoutput=1

\usepackage{xspace}
\usepackage{epsfig}
\usepackage{amssymb,amsmath}
\usepackage{cite}
\usepackage{url,booktabs}
\usepackage{color}
\usepackage{multicol}
\usepackage{multirow}
\usepackage{cancel}

\def\be{\begin{equation}}
\def\ee{\end{equation}}
\def\bea{\begin{eqnarray}}
\def\eea{\end{eqnarray}}
\def\beal{\begin{align}}
\def\eeal{\end{align}}

\def\be{\begin{equation}}
\def\ee{\end{equation}}
\def\bea{\begin{eqnarray}}
\def\eea{\end{eqnarray}}

\relax

\begin{document}
\topmargin-2.5cm

\begin{titlepage}
\begin{flushright}
\normalsize{  
FERMILAB-PUB-11-603-T \\
LPN11-60 \\
ZU-TH 22/11 
  }  
\end{flushright}
\vskip 0.3in
\begin{center}{\Large\bf
BMSSM Higgs Bosons at the 7 TeV LHC
}

\vskip .5in
{\bf Marcela Carena}$^{a,b,c}$,
{\bf Eduardo Pont\'{o}n}$^{d}$ {\bf and}
{\bf Jos\'e Zurita}$^{e}$
\vskip.5in

$^{a}$ 
{\em Theoretical Physics Department, Fermilab, Batavia, IL 60510, USA}\\
$^{b}$ 
{\em Enrico Fermi Institute and $^{c}$ Kavli Institute for Cosmological Physics, The University of Chicago, Chicago, IL 60637, USA}\\
$^{d}$ 
{\em Department of Physics, Columbia University,\\
538 W. 120th St, New York, NY 10027, USA}\\
$^{e}$ {\em Institut f{\"u}r Theoretische Physik, Universit\"at
Z\"urich, Winterthurerstrasse 190, CH-8057 Z\"urich, Switzerland.  }
\\
\end{center}
\vskip.5cm
\begin{center}
{\bf Abstract}
\end{center}
\begin{quote}

We consider the Higgs sector in extensions of the Minimal
Supersymmetric Standard Model by higher-dimension operators in the
superpotential and the K\"ahler potential, in the context of Higgs
searches at the LHC 7 TeV run.  Such an effective field theory (EFT)
approach, also referred to as BMSSM, allows for a model-independent
description that may correspond to the combined effects of additional
supersymmetric sectors, such as heavy singlets, triplets or gauge
bosons, in which the supersymmetry breaking mass splittings can be
treated as a perturbation.  We consider the current LHC dataset, based
on about $1-2~{\rm fb}^{-1}$ of data to set exclusion limits on a
large class of BMSSM models.  We also present projections for
integrated luminosities of 5 and 15~fb$^{-1}$, assuming that the ATLAS
and CMS collaborations will combine their results in each channel.
Our study shows that the majority of the parameter space will be
probed at the $2\sigma$ level with 15 fb$^{-1}$ of data.  A
non-observation of a Higgs boson with about 10 fb$^{-1}$ of data will
point towards a Higgs SUSY spectrum with intermediate $\tan \beta$ (
$\approx$ a few to~10) and a light SM-like Higgs with somewhat
enhanced couplings to bottom and tau pairs.  We define a number of
BMSSM benchmark scenarios and analyze the possible exclusion/discovery
channels and the projected required luminosity to probe them.  We also
discuss the results of the EFT framework for two specific models, one
with a singlet superfield and one with SU(2)$_L$ triplets.
\end{quote}
\vskip1.cm
\end{titlepage}
\setcounter{footnote}{0}
\setcounter{page}{1}
\newpage
\baselineskip18pt    

\noindent

\section{Introduction}
\label{sec:intro}

The search for a Standard Model (SM) Higgs boson responsible for
electroweak symmetry breaking has been the central focus of both the
Tevatron and the LHC in the recent past, and it remains one of the
main goals for the LHC in the years to come.  Present LEP, Tevatron
and LHC data have already placed strong direct bounds on the possible
mass of such a Higgs particle, leaving an allowed range between 114
GeV and 145 GeV, and above $\sim$ 450
GeV~\cite{ATLAS:HiggsCombined,CMS:HiggsCombined}.

Several shortcomings of the SM model (the Planck/weak scale hierarchy,
the origin of fermion masses and mixing angles, dark matter and
baryogenesis) could be addressed by Beyond the SM (BSM) extensions at
or somewhat above the TeV scale.  Some of these advocate a
perturbative extension as in supersymmetric theories, whereas others
involve strong dynamics as in extended
technicolor/topcolor/topcondensate theories or theories with extra
dimensions.  All these possible extensions address the question of
electroweak symmetry breaking via different mechanisms that may imply
the presence of various extended Higgs sectors, in which the Higgs
couplings to the known particles can vary significantly (or there may
be no Higgs at all).  It is of major importance to explore different
theoretical SM extensions that can alter the expected SM Higgs
production and decay modes, thereby allowing for very different
interpretations of the experimental Higgs mass bounds.

In the past years there has been extensive work in extensions of the
Higgs sector of the Minimal Supersymmetric Standard Model (MSSM) by
higher-dimension operators \cite{Strumia:1999jm}.  Indeed, several
different aspects related to fine-tuning, the Higgs potential, dark
matter, electroweak baryogenesis, flavor physics, and CP-violation
have been studied considering the effects of higher-dimension
operators of dimension five in the superpotential \cite{Batra:2008rc}
and dimension six in the K\"ahler
potential~\cite{Antoniadis:2007xc,Carena:2009gx,Carena:2010cs}.

In this article we will consider the Effective Field Theory (EFT)
approach described in Ref.  \cite{Carena:2009gx} and study the effects
of such a Beyond the MSSM (BMSSM) theory on Higgs searches at the LHC.
The EFT framework allows for a model-independent description of a
large class of extensions of the MSSM, which may include the combined
effects of many additional sectors at energies somewhat above the
electroweak scale, that can impact the Higgs phenomenology.  It
provides an opportunity to use the Higgs sector as a window on BMSSM
physics.  One the other hand, this EFT approach can also be
reinterpreted to explore the Higgs LHC potential for some specific
MSSM extensions, such as the addition of heavy singlets, triplets or
gauge bosons.  In particular, we will show that our approach can
reproduce to a good level of accuracy results of simple renormalizable
SUSY models.  One should note, however, that the study of the EFT in
superfield language relies on the assumption that the UV theory at a
scale M is supersymmetric up to small supersymmetry breaking effects
of order $m_s$ ($m_s$ being of order the electroweak scale), that can
be treated as a perturbation.

The phenomenology of the BMSSM Higgs sector up to dimension-six
operators in the superpotential and K\"ahler potential, including
all possible SUSY breaking effects via spurion superfields, was
studied in detail in Ref.~\cite{Carena:2009gx}.  
At leading order in $1/M$, the superpotential reads
\be\label{genw}
W=  \mu H_u H_d  + \frac{\omega_{1}}{2M}  [1+ \alpha_1 X] (H_u H_d)^2~,
\ee
where $H_u H_d = H_u^{0} H_d^{0} - H_u^+ H_d^-$, and $\omega_{1}$ and
$\alpha_1$ are dimensionless parameters that we assume to be of order
one.  The second term in the square brackets is the soft supersymmetry
breaking term parametrized via a (dimensionless) spurion superfield
$X=m_s \theta^2$.  At order $1/M^2$ there are no operators in the
superpotential, but several operators enter through the K\"ahler
potential:
\bea
\label{genk}
K&=&  H_d^{\dagger} \, e^{2V} H_d  + H_u^{\dagger} \, e^{2V} H_u  + \Delta K^{SUSY} + \Delta K^{\cancel{SUSY}}~,
\eea
where
\bea
\Delta K^{SUSY} &=& \frac{c_1}{2|M|^2} (H_d^{\dagger} e^{2V} H_d)^2   + \frac{c_2}{2|M|^2} (H_u^{\dagger} e^{2V} H_u)^2 + 
\frac{c_3}{|M|^2} (H_u^{\dagger} e^{2V} H_u) (H_d^{\dagger} e^{2V} H_d) 
\nonumber \\[0.5em]
&+ &  \frac{c_4}{|M|^2} |H_u H_d|^2 +  \left[\frac{c_6}{|M|^2} H_d^{\dagger} \, e^{2V} H_d   + 
\frac{c_7}{|M|^2} H_u^{\dagger} \, e^{2V} H_u \right] (H_u H_d)  + {\rm h.c.}~,
\label{eq:DeltaKSUSY}
\eea
and $\Delta K^{\cancel{SUSY}}$ contains all the SUSY-breaking
operators associated to the operators of Eq.~(\ref{eq:DeltaKSUSY}) by
multiplication by $X$, $X^\dagger$ or $X^\dagger X$.  We assume that
the coefficients of these SUSY-breaking operators are proportional to
the corresponding $c_{i}$, with proportionality constants of order
one, that we call $\beta_i$, $\gamma_i$ and $\delta_i$ (see
Ref.~\cite{Carena:2009gx} for the detailed definitions).

It was shown that the inclusion of the above higher-dimension
operators alleviates the tension present in the MSSM between the upper
theoretical bound of about 135 GeV and the non-observation at LEP of a
Higgs boson, as well as allowing for a Higgs phenomenology markedly
different from the MSSM. In Ref.~\cite{Carena:2010cs} we interpreted
the LEP and Tevatron Higgs boson bounds in the light of a parameter
scan of BMSSM models and defined a number of benchmark scenarios with
interesting Higgs phenomenology.  In this paper we study constraints
and prospects for detectability of extensions of the MSSM at the LHC
Run-I and we also present the results in specific models such as the
MSSM with an extra heavy singlet and the MSSM with extra heavy
triplets.  We base our results on the current data from Higgs searches
at the LHC at a center of mass energy of 7 TeV, with about $1-2~{\rm
fb}^{-1}$ of integrated luminosity per experiment (depending on the
channel)~\cite{ATLAS:htoWW2l2nu,Collaboration:2011as,Aad:2011uq,Collaboration:2011va,Aad:2011ec,
CMS:hWW ,Collaboration:2011ww ,CMS:Htogaganote,
CMS:htoZZ,CMS:taus,ATLAS:taus,CMS:chargedHiggs,CMS:VHtobb}, and
extrapolate the expected results for two scenarios: 5 and 15 fb$^{-1}$
of integrated luminosity.

The most sensitive channels to search for a SM Higgs boson at the LHC
are highly dependent on its mass.  If the Higgs is light, in the
$115-120$ GeV range, the most sensitive channel is the diphoton
($\gamma \gamma$) one.  For intermediate masses ($120-205$ GeV) it is
the WW channel, and for heavier bosons ($205-600$ GeV), the ZZ
channel.  When considering neutral Higgs bosons of an extended sector
(like in supersymmetric theories), generically denoted by $\Phi$, this
situation can change according to how these Higgs bosons couple to the
SM particles.  Nevertheless, we expect at least one of them to couple
sizably to the $W$ and $Z$ gauge bosons, and therefore these decay
modes may also be useful to probe neutral Higgs bosons from such
non-minimal sectors.  Similarly, sizable couplings to the top quark
can induce important couplings to photons at loop level, just like for
the SM Higgs.  In addition, Higgs decays into down-type fermions, such
as $V\Phi, \Phi \to b \bar{b}$, $qq\Phi, \Phi \to \tau^+\tau^-$, or
the inclusive decay into $\tau^+ \tau^-$ can also be useful to probe a
neutral Higgs boson.  In the SM, the latter channel does not
constitute an early discovery mode.  However, if one has an enhanced
coupling to bottom pairs that enhances the production cross section,
as in the large $\tan \beta$ regime of the MSSM, it can turn into a
discovery mode (see Ref.~\cite{Carena:2011fc} for a recent study of
the LHC-Run~I reach within the MSSM).

We are particularly interested in the first few to ten inverse
femtobarns of LHC data, where many of these channels will start to
show sensitivity to the Higgs boson.  Other implications of early LHC
results have been recently considered in the context of SUSY singlet
extensions~\cite{Ellwanger:2011mu}, in more general 2HDM
scenarios~\cite{Chang:2011kr}, and also for dark Higgs models~\cite{Weihs:2011wp} (where the SM Higgs sector is enlarged with a SM singlet). With the current dataset, Tevatron
bounds coming from the WW decay mode
\cite{Aaltonen:2010sv,Aaltonen:2010yv} are already superseded by the
LHC \cite{ATLAS:htoWW2l2nu,Collaboration:2011as,CMS:hWW}.  With a few
inverse femtobarn of data, the diphoton channel will be able to probe
points where the cross section times branching fraction is close to
the SM one \cite{CMS:HiggsNote,Atlas:HiggsNote}, which may be compared
to the current Tevatron factor of about 15
\cite{CDF:htogammagamma,D0:htogammagamma}.  In the low mass region,
$V\Phi, \Phi \to b \bar{b}$ rates close to the SM one can be explored
at the Tevatron, whereas they require a larger dataset at the LHC. The
$ZZ$ channel\cite{Atlas:HiggsNote} will also be effective to rule out
Higgs bosons with a mass in the $200-600$ GeV range.  Such a mass
range cannot be probed at the Tevatron since the corresponding
production cross sections are very small.
 
This paper is organized as follows.  In Section~\ref{sec:Hprobe}, we
summarize the most relevant Higgs search channels at the LHC Run-I. In
Section~\ref{sec:results} we present our results, showing the reach of
LHC Run-I for different  BMSSM scenarios and specific decay channels. We also
update the prospects for the benchmark points presented in
Ref.~\cite{Carena:2010cs}, separating the analysis into the low and
large $\tan\beta$ regimes.  In section 4 we study MSSM extensions with
a heavy singlet and with heavy triplets using the EFT approach.  We
conclude in Section~\ref{sec:conclu}.

\section{Probing the Higgs sector with early LHC data}
\label{sec:Hprobe}

In this section we present all the Higgs search channels at the LHC
that will be used in our study.  We will employ data when available in
order to take into account the most up-to-date experimental details
(e.g. efficiencies, acceptances and background estimations) from the
LHC collaborations.  However, for $V\Phi,\Phi \to b\bar{b}$ and
$qq\Phi, \Phi \to \tau^+\tau^-$ we employ the available MC
simulations.\footnote{These channels do not yet play a major role, but
we have included them for completeness.} Throughout this work we will
consider the following Higgs search channels:
\begin{multicols}{2}
\begin{itemize}
\item[a)] $pp \to \Phi \to WW, ZZ, \gamma \gamma$~,
\item[b)] $pp \to \Phi \to \tau^{+} \tau^{-}$~,
\item[c)] $V\Phi, \Phi \to b \bar{b}$, and $qq\Phi, \Phi \to \tau^{+} \tau^{-}$~,
\item[d)] $t \to H^+ b$~.
\end{itemize}
\end{multicols}
The channels listed in a) and c) are conventional SM Higgs search
channels at the LHC, the only difference being that c) are not expected to be
discovery modes for a SM Higgs boson in LHC run I.  The inclusive tau channel, b),
has a very low rate in the SM, but can be enhanced in non-minimal
scenarios.  This situation arises when one neutral Higgs boson couples
very weakly to gauge bosons and has enhanced couplings to down-type
fermions, like in the large $\tan \beta$ limit of the MSSM, or in the
decoupling limit, where $m_A \sim m_H \gg m_{h}$ and H is gaugephobic.
Finally, a charged Higgs lighter than the top quark can also be looked
for in channel d), and can be interesting at the LHC with a relatively
small dataset.

We will summarize first the ``SM-like Higgs searches'', focusing on
the channels a), but also including channels c).  Later, in Subsection
\ref{subsec:htotaus}, we will analyze separately the inclusive tau
case, b).  Finally we will discuss the reach for a charged Higgs boson
in Subsection \ref{subsec:hplus}.

\subsection{SM-like Higgs Searches}
\label{subsec:Hother}

%
\begin{table}[t]
\begin{center}
\begin{tabular}{|c|cc|c|c|c|}
\hline
\rule{0mm}{5mm}
\multirow{2}{*}{Channel} & 
\multicolumn{2}{|c|}{Lum. (fb$^{-1}$)} & 
\multirow{2}{*}{} Mass range & 
\multirow{2}{*}{Ref.} \\
 & ATLAS & CMS  & (GeV) & 
 \\ [0.3em]
\hline
\rule{0mm}{5mm}
$p p \to \Phi \to WW$ & 1.7  &   1.55 & 115-600 &\cite{ATLAS:htoWW2l2nu,Collaboration:2011as,CMS:hWW}
\\ [0.3em]
\hline 
\rule{0mm}{5mm}
$p p \to \Phi \to ZZ$ & $1.04$
--$2.28$ & $1.1$
--$1.7$ & 120-600 &  \cite{Aad:2011uq,Collaboration:2011va,Aad:2011ec,CMS:htoZZ} \\ [0.3em]
\hline
\rule{0mm}{5mm}
$p p \to \Phi \to \gamma \gamma$  & 1.08 & 1.7 & 110-150 &  \cite{Collaboration:2011ww,CMS:Htogaganote} \\ [0.3em]
\hline
\rule{0mm}{5mm}
$p p\to \Phi \to \tau^+ \tau^-$ & $1.06$  & 1.6 & 90-600 & 
\cite{ATLAS:taus,CMS:taus}
\\ [0.3em]
\hline
\rule{0mm}{5mm}
$V\Phi, \Phi \to b \bar{b}$ & $-$  & $ 1.1$  & 110-135 & 
\cite{CMS:VHtobb} \\ [0.3em]
\hline
\rule{0mm}{5mm}
$qq\Phi, \Phi \to \tau^+ \tau^-$ & 1  & $-$ & 110-130 & 
\cite{Atlas:HiggsNote} \\ [0.3em]
\hline
\rule{0mm}{5mm}
$t \to H^+ b, H^+ \to \tau^+ \nu_\tau$ & $-$ & 1.1 & 80 - 160 & \cite{CMS:chargedHiggs} \\ [0.3em]
\hline
\end{tabular}
\end{center}
\caption{List of LHC channels used in this study, indicating the
luminosity used by the collaborations in the analysis.  The ``$-$''
indicates that no data for that particular channel has been presented
by the corresponding collaboration.  Here, $\Phi$ stands for any
neutral Higgs boson.  The production mechanisms considered are
gluon-fusion, vector-boson fusion, associated production with $Z,W, t
\bar{t}$ and also $b \bar{b} \to \Phi$ (only relevant for large $\tan
\beta)$. }
\label{tab:LHC_chan}
\end{table}
In our analysis we consider all present direct experimental bounds on
a SM Higgs and reinterpret them in terms of our BMSSM scenarios.  We
take into account all the bounds from LEP and Tevatron searches via
HiggsBounds~v2.1.1\cite{Bechtle:2008jh,Bechtle:2011sb}, and focus on
those models that are \textit{not excluded} by these experiments at
the 95\% CL. For the LHC analysis, we consider the most recent data of
the ATLAS and CMS collaborations in the $WW$
\cite{ATLAS:htoWW2l2nu,Collaboration:2011as,CMS:hWW}, $ZZ$
\cite{Aad:2011uq,Collaboration:2011va,Aad:2011ec,CMS:htoZZ} and
$\gamma \gamma$ \cite{Collaboration:2011ww,CMS:Htogaganote} channels,
based on $1.04-2.28~{\rm fb}^{-1}$ of integrated luminosity.  These
channels are the most sensitive ones to a SM Higgs boson, in the mass
ranges given earlier.  For the associated production with a weak gauge
boson (Higgs-strahlung), with the Higgs decaying into $b \bar{b}$,
both ATLAS~\cite{Atlas:VHtobb} and CMS~\cite{CMS:VHtobb} have recently
presented results with about $1~{\rm fb}^{-1}$ of data.
The current CMS search is sensitive to a rate of about 6 times the SM
one, while the ATLAS search can only exclude a rate of about 20 times
the SM. Note, however, that in the MonteCarlo 2010
ATLAS sample~\cite{Atlas:HiggsNote}, where the boosted $b\bar{b}$ pair
techniques of~\cite{Butterworth:2008iy} were employed, the expected
sensitivity with 1 fb$^{-1}$ was very similar to the recent CMS
result.  For the $qq\Phi$, $\Phi \to \tau^+\tau^-$ channel, for which
no LHC collaboration has presented data, we employ the MC2010 sample.
Finally, we add the LHC searches for $\Phi \to \tau^+ \tau^-$, which
are discussed in detail in Section~\ref{subsec:htotaus}.

In the channels where both CMS and ATLAS have presented data, we
combine their results following the simple prescription described in
Refs.~\cite{Draper:2009fh,Draper:2009au}.  Although this procedure may
be overly simplistic, and a careful combination by the experimental
collaborations would be most welcome, it allows us to get an idea of
the present exclusion bounds with the information available.
Therefore, when applied to our scan over BMSSM scenarios, our
\textit{current exclusion} statements will refer to such a combination
of the observed limits by both experiment.  For channels where only
one collaboration has presented an analysis, we will base our current
exclusion on that analysis.  The \emph{projections} of the LHC reach
for a given luminosity, on the other hand, are computed using the
method described in Appendix \ref{app:significances}.  For channels
where only one collaboration has presented data, we assume that the
expected limit of the other collaboration will be similar, and
``double'' the expected projected dataset (we call it CMS $\times$ 2
or ATLAS $\times$ 2).\footnote{Given that, as explained above, the
expected ATLAS sensitivity in the $V\Phi, \Phi \to b \bar{b}$ channel
after improving their analysis should be similar to the expected CMS
exclusion limit, we employ the CMS x 2 prescription for the
projections in this channel.  } Thus, our \textit{projections} should
always be interpreted as what would be expected from a combination of
both experiments.  A summary of the various datasets used in this work
is presented in Table~\ref{tab:LHC_chan}.

\begin{figure}[t]
\begin{center}
\begin{minipage}[b]{0.45 \linewidth}
\begin{center}
\includegraphics[width=1\textwidth]{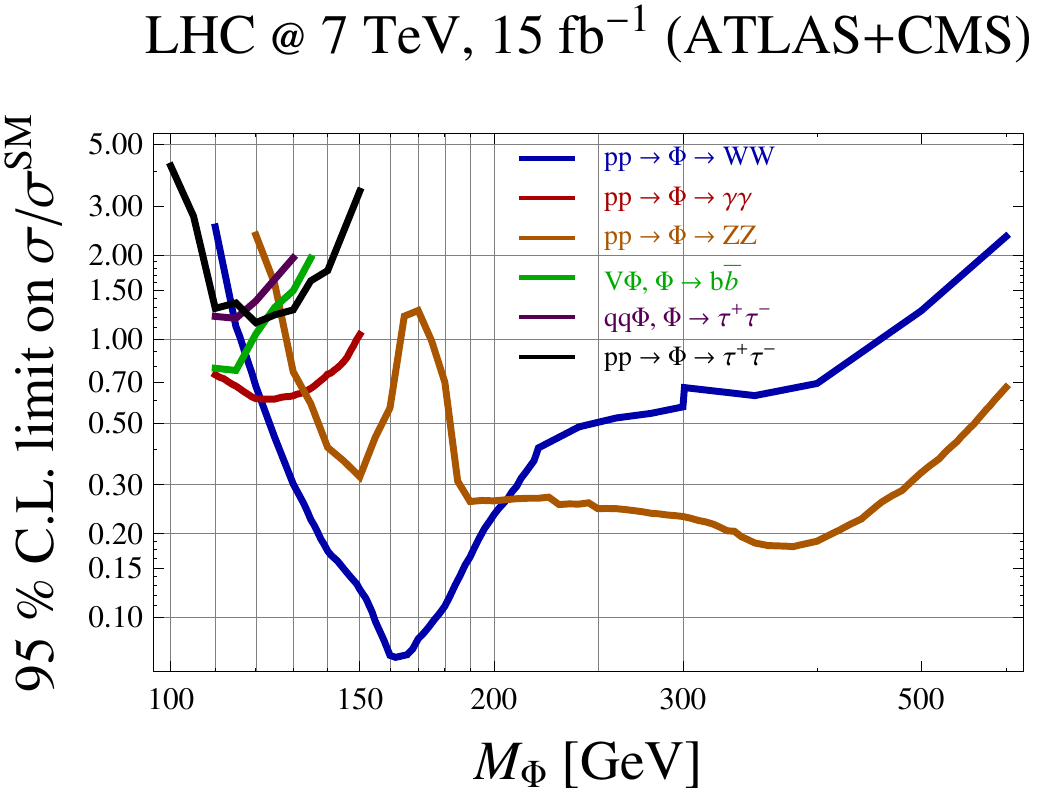}
\newline
(a)
\end{center}
\end{minipage}
\hspace{0.5cm}
\begin{minipage}[b]{0.45\linewidth}
\begin{center}
\includegraphics[width=1\textwidth]{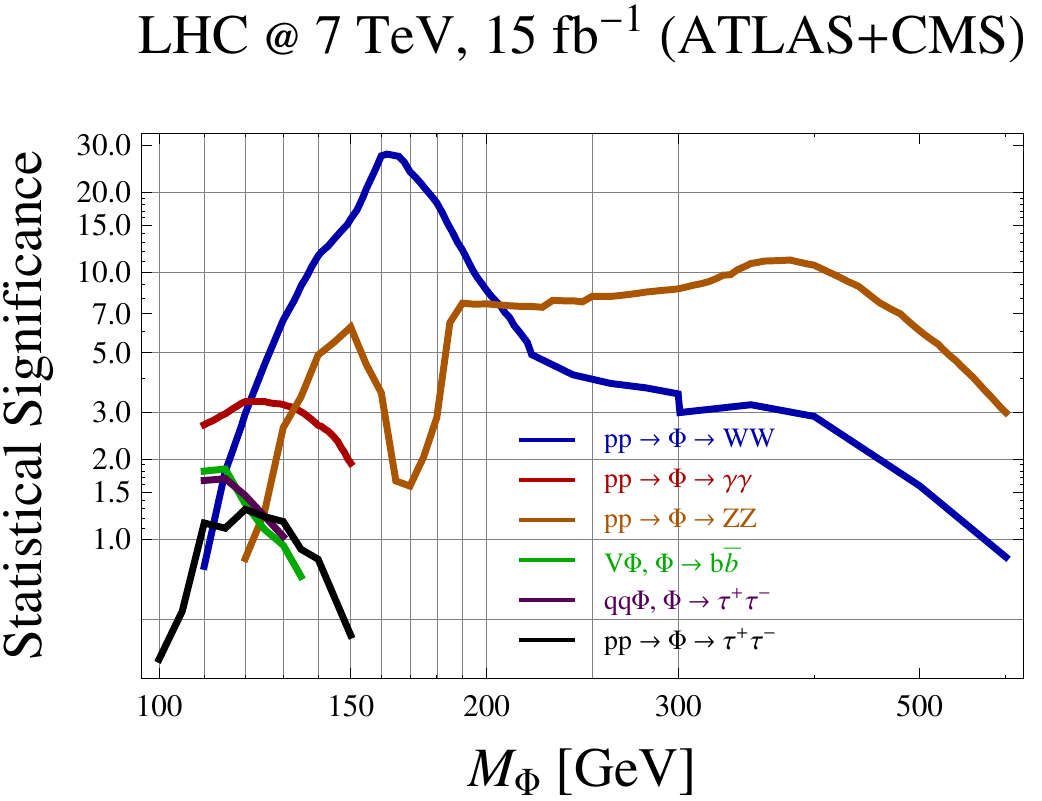}
\newline
(b)
\end{center}
\end{minipage}
\caption{ LHC reach (a) and significances (b) for the SM Higgs boson
with $15~\rm{fb}^{-1}$, combining both experiments.  The color coding
is as follows: WW (blue), ZZ (orange), $\gamma \gamma$ (red), $\tau^+
\tau^-$ (black), $V\Phi, \Phi \to b \bar{b}$ (green), $qq\Phi, \Phi
\to \tau^+ \tau^-$ (purple) and $\tau^+ \tau^-$ inclusive (black). Figure taken from Ref.~\cite{Weihs:2011wp}.
}\label{fig:lhc_reach}
\end{center}
\end{figure}

We present in Fig.~\ref{fig:lhc_reach} (taken from
Ref.~\cite{Weihs:2011wp}) the expected 95 $\%$ C.L. exclusion limit on
Higgs production cross sections (including the corresponding decay
rates) normalized to their SM values as a function of the Higgs mass
for a total integrated luminosity of 15 fb$^{-1}$.  We also show the
statistical significance of the different channels as a function of
the Higgs mass for the same integrated luminosity.  We see that the
$WW$ channel has exclusion power down to $M_{\Phi} \sim$115 GeV and up
to $M_{\Phi} \sim$450 GeV. It is actually the most sensitive channel
for $120~{\rm GeV} \lesssim M_{\Phi} \lesssim 205~{\rm GeV}$, while
for larger masses the ZZ signal takes over this role.  For masses in
the 115-150 GeV range, one can test signals in the diphoton channel as
low as 0.6-0.7 times the SM rate (2$\sigma$ exclusion), and test the
SM up to about $3 \sigma$.  The remaining channels are less powerful
in probing the SM Higgs at the LHC Run-I.

\subsection{Non-SM Neutral Higgs searches in the $\tau^+ \tau^-$ channel}
\label{subsec:htotaus}

This search is important for neutral Higgs bosons in the MSSM at large
values of $\tan \beta$, where the bottom-Yukawa coupling is enhanced,
thus yielding a significant increase in the rate.  The current analyses of
ATLAS~\cite{ATLAS:taus} ($1.06~\rm{fb}^{-1}$) and CMS~\cite{CMS:taus}
($1.6~\rm{fb}^{-1}$), taken individually, are able to probe a rate of
about 10 times the SM one, for masses between 110-150 GeV, already one
order of magnitude better than the current results from Tevatron
(combining CDF and D0)\cite{Benjamin:2010xb}.

In this study, we are interested in the bounds on the $h/H/A \to
\tau^+\tau^-$ cross sections presented by the LHC collaborations,
which extend up to Higgs masses of about $600~{\rm GeV}$.
ATLAS~\cite{ATLAS:taus} reports individual 95\% CL limits for the $gg
\to \Phi$ and $b \bar{b} \Phi$ production modes ($\Phi = h,H,A$),
while CMS~\cite{CMS:taus} presents a combined result of these two
production channels.  In order to obtain the exclusion limit, we
compute in each of our model points the $gg \to \Phi \to \tau^+
\tau^-$ and $bb\Phi, \Phi \to \tau^+ \tau^-$ rates, and derive the Q
values, as defined in Appendix~\ref{app:significances} [see
Eq.~(\ref{eq:defQ})], one for each of the three experimental limits
above.  The production cross sections and branching fractions at the
LHC are taken from
Ref.~\cite{LHCHiggsCrossSectionWorkingGroup:2011ti}, except for the
$bb \Phi$ cross section, that was obtained using the code bbh@NNLO
\cite{Harlander:2003ai} with the MSTW 2008 PDF set
\cite{Martin:2009iq}.  If the masses of two or more Higgs bosons fall
within $10~{\rm GeV}$ of each other we add the corresponding signals.
We then combine the three significances (from ATLAS in $gg \to \Phi$,
ATLAS in $bb\Phi$ and CMS combined) in quadrature to obtain a total
significance in the $\tau^+\tau^-$ channel for each scenario of our
BMSSM parameter scan.

\subsection{Charged Higgs searches in top decays}
\label{subsec:hplus}

Besides the neutral Higgs sector, one can also probe at the LHC a
charged Higgs boson, produced in the decay of the top quark.  This
decay mode is effective only for $m_{H^+} < m_{t} - m_{b}$.  The
tree-level partial decay widths for $t \to W^+ b$ and $t \to H^+ b$
are given by \cite{Gunion:1989we}
\be\label{eq:toptoWb}
\Gamma (t \to W^+ b) = \frac{G_F}{8 \pi \sqrt{2}} m_t^3 \lambda^{1/2} (1,x_b,x_w) 
\left[x_W (1+x_b) + (1-x_b)^2 -2 x_W^2 \right] \, ,
\ee
and
\be\label{eq:toptoHb}
\Gamma (t \to H^+ b) = \frac{G_F}{8 \pi \sqrt{2}} m_t^3 \lambda^{1/2} (1,x_b,x_{H^+}) 
\left[ \left( \frac{1}{\tan \beta^2} + x_b \tan \beta^2 \right) (1+x_b-x_{H^+}) + 4 x_b \right] \, , 
\ee
where $\lambda(a,b,c)=a^2+b^2+c^2-2ab-2ac-2bc$ and $x_i = m_i^2/
m_t^2$.  We also implement the NLO QCD corrections to both $t \to W^+
b$ \cite{Jezabek:1988iv} and $t \to H^{+}
b$~\cite{Li:1990cp,Czarnecki:1992zm}.  For the $W$-channel we use a
QCD K-factor $K = 1 - (2\alpha_{s}/3\pi)(2\pi^2/3 - 5/2) \approx
0.91$~\cite{Jezabek:1988iv}, where $\alpha_{s} \approx 0.107$ is
evaluated at $\mu = m_{t}$.  For the Higgs channel we implement the
results of \cite{Czarnecki:1992zm}, which hold for any value of
$\tan\beta$.  At small $\tan\beta$ these QCD corrections are small
(below 10\%), but they can be sizable at larger $\tan\beta$.  Also, at
large $\tan\beta$ the SUSY QCD corrections can be important.  To take
these latter effects into account we use the SUSY QCD corrections
presented in~\cite{Carena:1999py}, which amount to using an effective
bottom Yukawa coupling corrected by $1/(1+\Delta_{b})$, where
$\Delta_{b}$ depends on the SUSY spectrum~\cite{Deltamb}.

\begin{figure}[t]
\begin{center}
\begin{minipage}[b]{0.45\linewidth}
\begin{center}
\includegraphics[width=1\textwidth]{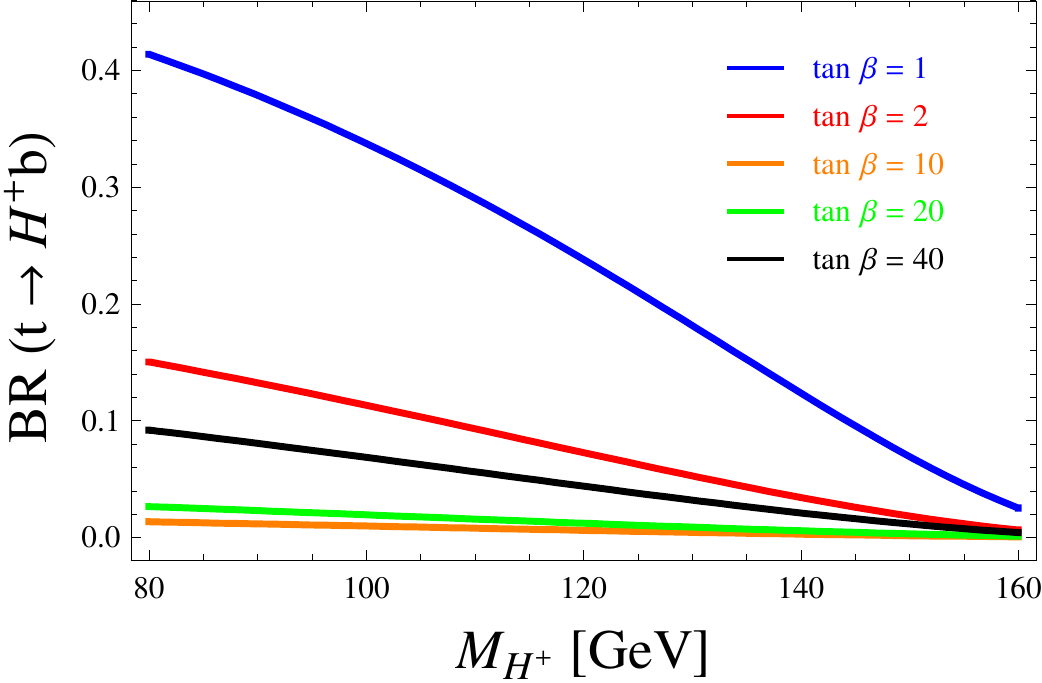}
\newline
(a)
\end{center}
\end{minipage}
\hspace{0.5cm}
\begin{minipage}[b]{0.45\linewidth}
\begin{center}
\includegraphics[width=1\textwidth]{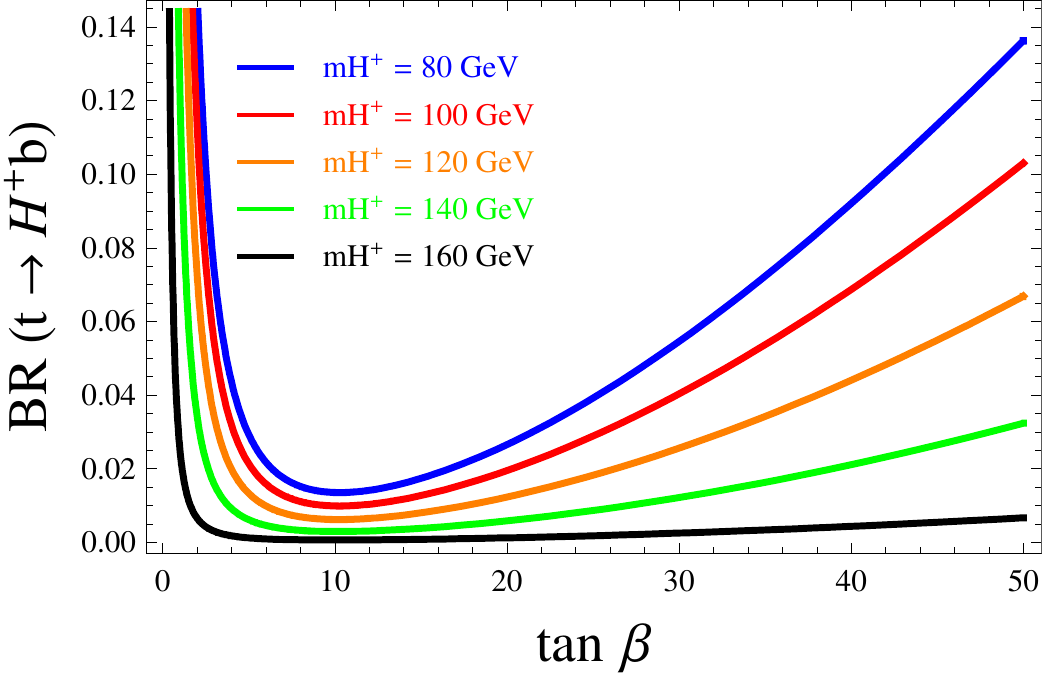}
\newline
(b)
\end{center}
\end{minipage}
\caption{ Branching fraction of the top quark into a charged Higgs and
a bottom quark as a function of (a) $m_{H^+}$ and (b) $\tan \beta$. 
SUSY QCD corrections are not included.}
\label{fig:toptoHb}
\end{center}
\end{figure}
As a guide, in Fig.~\ref{fig:toptoHb} we show the branching fraction
of the top quark decaying into a charged Higgs plus a bottom quark, as
a function of the charged Higgs mass (left panel) and also as a
function of $\tan \beta$, including only the QCD corrections
of~\cite{Czarnecki:1992zm}.  However, when applied to our scan over
BMSSM scenarios, and for the $\tan\beta > 10$ cases, we will also
include the SUSY QCD corrections described above (assuming gluinos and
squarks at 1 TeV, negligible $A$-terms and $\mu > 0$~\footnote{For
$\mu < 0$ the corrections due $\Delta_{b}$ can significantly enhance
the branching fraction into the charged Higgs
channel~\cite{Carena:1999py}.}).  For the values of $\tan \beta$ that
will be employed in this work (2 and 20), we see that this branching
ratio is always below 20\%, and decreases with increasing $m_{H^+}$,
due to a phase space suppression.  From the right panel we see that
there is a minimum around $\tan \beta \sim \sqrt{m_{t}/m_b} \sim 8$,
where $m_{b}$ is evaluated at the scale of the top mass.  For a fixed
charged Higgs mass, the branching fraction grows for either very large
or small values of $\tan \beta$.

\begin{figure}[t]
\begin{center}
\includegraphics[width=0.45\textwidth]{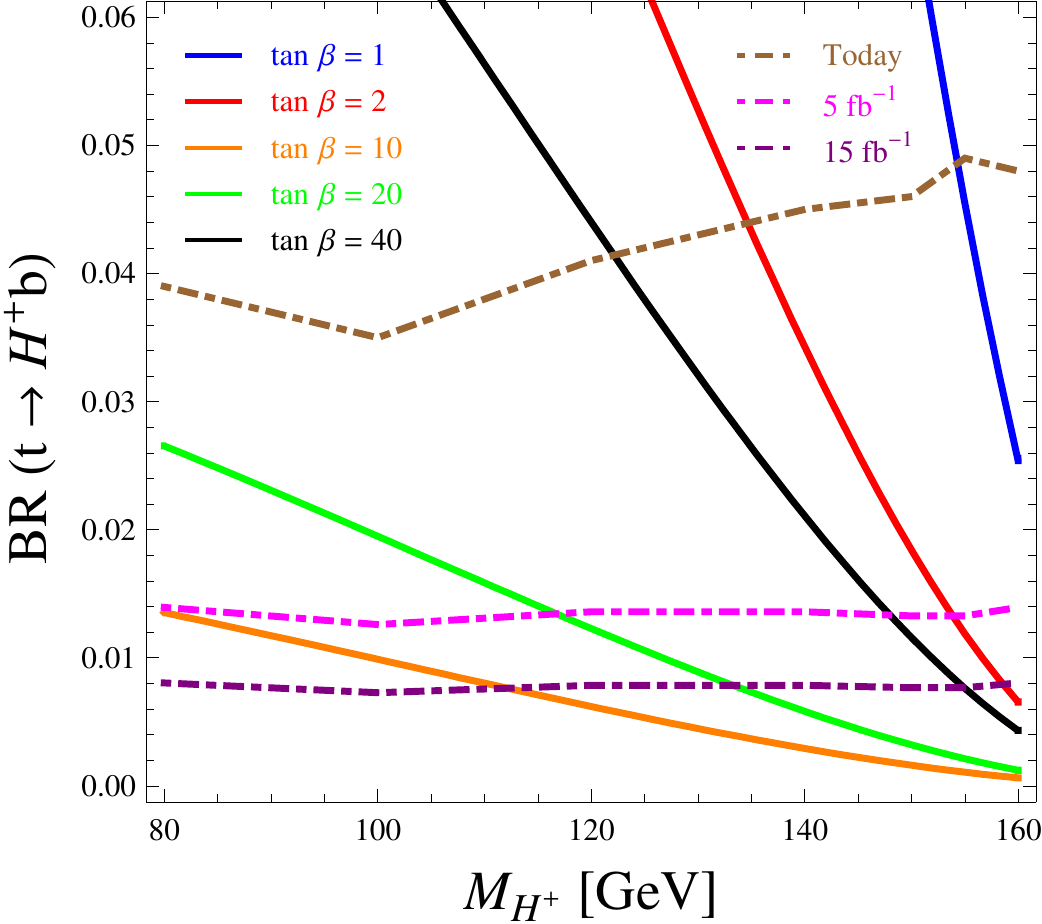}
\end{center}
\caption{The current CMS 95\% CL upper bound on ${\rm BR}(t \to H^+b)$
(dot-dashed brown)~\cite{CMS:chargedHiggs}, together with the
projected (CMS $\times$2) LHC reach for 5 and 15 fb$^{-1}$, shown in
dot-dashed purple and pink curves.  The region above the dot-dashed
curves would be excluded at the specified total integrated luminosity.
The solid lines correspond to $\tan \beta$= 1 (blue), 2 (red), 10
(orange), 20 (green) and 40 (black).  SUSY QCD corrections are not
included.}
\label{fig:toptoHbCMS}
\end{figure}
To set limits and make projections, we use the latest CMS search
\cite{CMS:chargedHiggs} for a charged Higgs produced in top decays,
which assumes that ${\rm BR}(H^{+} \to \tau^{+} \nu_\tau) = 1$.  We
will therefore apply this limit only to models where ${\rm BR}(H^+ \to
\tau^{+} \nu_{\tau}) > 0.9$, and in those cases we will interpret the
CMS bound as applying to ${\rm BR} (t \to H^+ b) \times {\rm BR}(H^+
\to \tau \nu)$.  We expect that this procedure will give a good
estimate of the LHC charged Higgs reach for BMSSM scenarios with such
a dominant $\tau \nu$ decay channel.\footnote{ We also ignore the
modified $H^+ b t$ coupling due to the canonical renormalization
required when introducing dimension-six operators (see
Ref.~\cite{Carena:2009gx}).  This is a small effect, always below 4
(3) \% for $\tan \beta =2$ (20), and thus will be neglected throughout
this paper.} In Fig.~\ref{fig:toptoHbCMS}, the dot-dashed brown curve
corresponds to the CMS observed limit on ${\rm BR}(t \to H^+ b)$.  We
also show curves of ${\rm BR}(t \to H^+ b)$ as a function of
$m_{H^{+}}$ for fixed $\tan\beta$, using Eqs.~(\ref{eq:toptoWb}) and
(\ref{eq:toptoHb}), with the QCD corrections
of~\cite{Czarnecki:1992zm} taken into account.

We see from these figures that for $\tan \beta=2$ the charged Higgs is
very constrained: one can exclude values below about 135 (160) GeV
with 1 (15) fb$^{-1}$, where for the projection we use the ``CMS
$\times$ 2'' prescription.  For $\tan \beta=20$, where the NLO QCD
corrections are larger, there is currently no exclusion for $m_{H^+} >
80$ GeV. With 5 (15) fb$^{-1}$ one will start to probe masses up to
about 120 (135) GeV. However, we note that the inclusion of the
SUSY-QCD corrections for the supersymmetric parameters considered in
this work weakens the 7 TeV LHC run reach, and can only probe charged
Higgs masses up to 105 (125) GeV for total integrated luminosities of
5 (15) fb$^{-1}$.  On the other hand, for other choices of the
supersymmetric parameters (e.g.~with $\mu < 0$), larger charged Higgs
masses could be probed for such luminosities.

\section{Results}
\label{sec:results}

In this Section we present the results of our analysis for BMSSM
scenarios.  We use the same sample of points that was used in
Ref.~\cite{Carena:2009gx,Carena:2010cs}, to which we added a sparse scan over $\tan \beta$ (see Ref.~\cite{Carena:2009gx}
for technical details on how the scan was performed).  In brief, this
points are consistent with electroweak precision data and do not
receive sizable corrections from higher-dimension operators (namely,
the perturbative series in powers of $1/M$ seems to converge).
Moreover, our working assumption is that there exist BMSSM degrees of
freedom with masses of about $1~{\rm TeV}$ that couple with order-one
couplings to the MSSM Higgs sector.  Thus, these scenarios always
represent significant departures from the MSSM and our conclusions
regarding exclusion or discovery prospects cannot be simply applied to
the MSSM limit.  We have considered stops of about 300 GeV to
emphasize that radiative corrections play a minor role in the Higgs
spectrum, but they could be somewhat heavier without significantly
changing our results.\footnote{ A recent ATLAS study shows some
sensitivity to a lightest stop with mass $m_{\tilde{t}_{1}} \sim
300~{\rm GeV}$ provided the gluino mass is around $500~{\rm
GeV}$~\cite{Atlas:StopsNote}.  Similarly, sbottoms can be bounded by
about $600~{\rm GeV}$ provided the gluino mass is below $750~{\rm
GeV}$~\cite{Atlas:SbottomsNote}, and by about $250~{\rm GeV}$ for
neutralino masses below $110~{\rm GeV}$~\cite{Abazov:2010wq}.
However, for heavier gluinos or neutralinos the bounds on the
stop/sbottom masses essentially disappear.  Furthermore, gluinos and
first two generation squarks may have to be heavier than about $1~{\rm
TeV}$~\cite{Chatrchyan:2011ek}.  These latter particles play no
relevant role in our study.} For the majority of our analysis we
explore two values for $\tan \beta$: 2 and 20, which are taken to be
representative of the small and large $\tan \beta$ regimes.  At even
larger values of $\tan\beta$, the effects of the BMSSM sector are
smaller.  We also consider a smaller parameter scan for intermediate
values of $\tan \beta$ between 4 and 8, which turns to be more
challenging for the ongoing LHC run.

We will present the current and projected LHC Run I constraints on our
sample, and update the benchmark scenarios presented in
Ref.~\cite{Carena:2010cs}.  As we will see some of these scenarios
have been excluded by the latest LHC studies.  We will also present
further benchmark points that illustrate the exclusion/discovery
prospects at the 7 TeV LHC run.

\subsection{Global constraints from the LHC}

\begin{figure}[h]
\begin{center}
\begin{minipage}[b]{0.45\linewidth}
\begin{center}
\includegraphics[width=\textwidth]{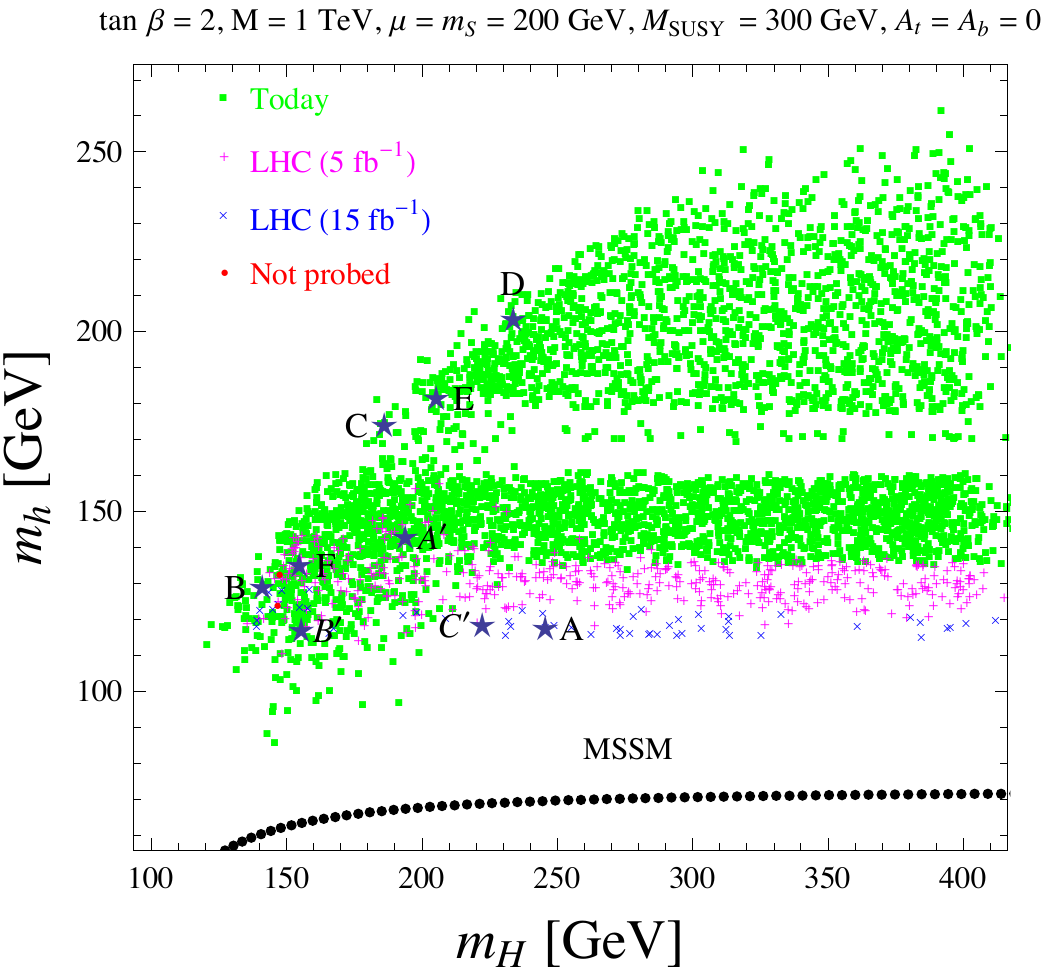}
\newline
\end{center}
\end{minipage}
\hspace{0.5cm}
\begin{minipage}[b]{0.45\linewidth}
\begin{center}
\includegraphics[width=\textwidth]{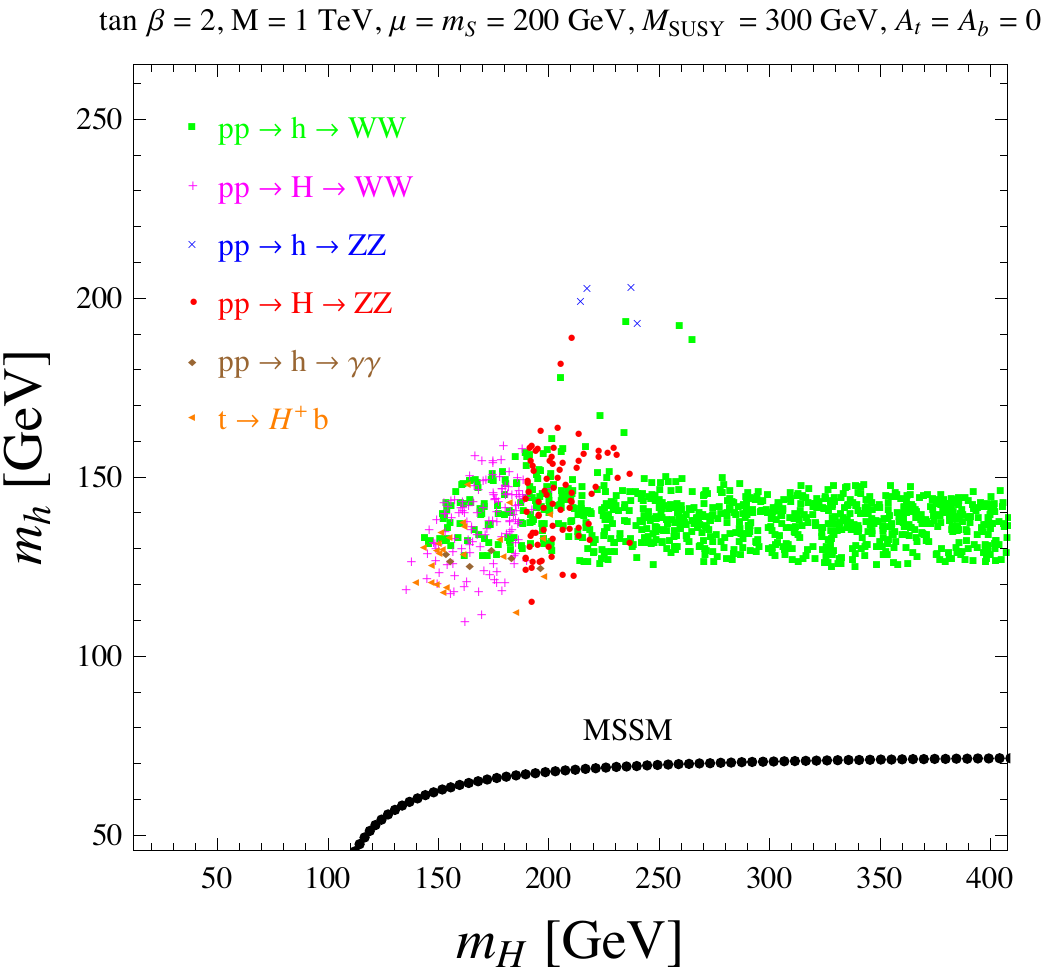}
\newline
\end{center}
\end{minipage}
\caption{\small Scan over BMSSM scenarios, for $\tan\beta=2$.  Upper:
Points not excluded by LEP or Tevatron data at the 95\% CL. We show
points currently excluded by the LHC with 1 fb$^{-1}$ in green.  The
LHC exclusion reach for 5 and 15 fb$^{-1}$ is shown in magenta and
blue, while points outside the LHC-Run I reach are plotted in red.
Lower: Models that can be discovered (5$\sigma)$ after 15 fb$^{-1}$ of
collected data, assuming an ATLAS/CMS combination in each separate
channel.  We include here those points that are currently excluded at
less than $3\sigma$ (most currently excluded points are excluded at
more than $3\sigma$).  We indicate the discovery mode: $pp \to h \to
WW$ (green), $pp \to H \to WW$ (magenta), $pp \to h \to ZZ$ (blue),
$pp \to H \to ZZ$ (red), $pp \to h \to \gamma \gamma$ (brown) and $t
\to H^+b$ (orange).  Note that the color code is different in the two
plots.  }
\label{fig:mavsmltoday}
\end{center}
\end{figure}

In this subsection we present plots that illustrate some generic
features of BMSSM scenarios in connection to LHC Higgs phenomenology.
In Fig.~\ref{fig:mavsmltoday} we show our scan of points, in the
$m_H-m_h$ plane, for $\tan \beta=2$.  In the left panel we plot all
the points currently \textit{not excluded} by LEP or Tevatron data.
We show in green those points that are excluded by the most recent LHC
limits (combining data from both collaborations), while those that
require a total integrated luminosity of 5 and 15 fb$^{-1}$ to be
within reach of the LHC, are shown in magenta and blue, respectively.
Points outside of LHC Run-I reach are shown in red.  The dotted line
shows the MSSM result, which we provide as a reference (for such light
stops, the MSSM would be excluded by LEP).  
The LEP constraints rule out points with low values of $m_h$.  The few
allowed points with $m_{h} \lesssim 100~{\rm GeV}$ are not probed by
LEP due to a very reduced coupling of the lightest CP-even Higgs to
gauge bosons (H is SM-like), which suppresses the Higgs-strahlung
production cross section.  Tevatron bounds explain the absence of
points with $m_h$ in the 160-180~GeV range, the exceptions
corresponding to cases where the coupling of $h$ to WW and ZZ is
sufficiently smaller than their SM counterpart.  The current LHC
limits extends this range to 135-250 GeV. With 15 fb$^{-1}$, one will
further probe points down to 115 GeV, either in the $h \to WW$ or $h
\to \gamma \gamma$ channel, thus excluding most of the scanned points.
The cases that cannot be tested at the 2$\sigma$ level with
15~\rm{fb}$^{-1}$ correspond to points where the CP-odd Higgs A is
light and therefore one or both CP-even Higgs bosons have a sizable
branching ratio into $AA$, or points where the branching ratio into $b
\bar{b}$ is enhanced compared to the SM, thus reducing the $WW$ and
$\gamma \gamma$ branching fractions.

In the right panel of Figure~\ref{fig:mavsmltoday} we plot those
points which are within the \textit{discovery} reach of the LHC with
15 fb$^{-1}$ of total integrated luminosity, combining both
experiments.  Besides the currently allowed points (significance less
than $2\sigma$), we include in this plot also those points that are
currently excluded at a significance between $2\sigma$ and $3\sigma$,
to account for a possible downward fluctuation in current data.  For
this subset (i.e.~``exclusion significance below $3\sigma$''), we
indicate by the color code the most sensitive channel to discover a
Higgs boson: $pp \to h \to WW$ (green), $pp \to H \to WW$ (magenta),
$pp \to h \to ZZ$ (blue), $pp \to H \to ZZ$ (red), $pp \to h \to
\gamma \gamma$ (brown), and $t \to H^+b$ (orange).

\begin{figure}[h]
\begin{center}
\begin{minipage}[b]{0.45\linewidth}
\begin{center}
\includegraphics[width=1\textwidth]{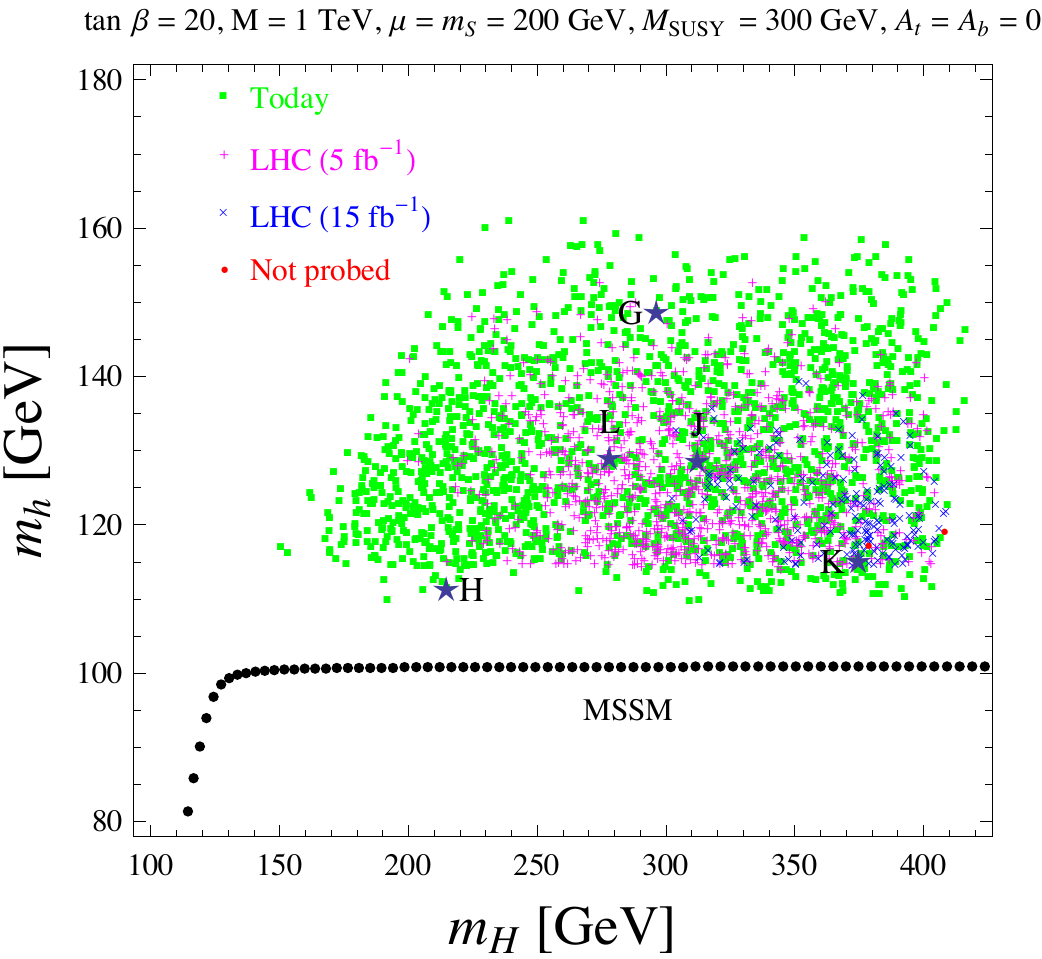}
\newline
\end{center}
\end{minipage}
\hspace{0.5cm}
\begin{minipage}[b]{0.45\linewidth}
\begin{center}
\includegraphics[width=1\textwidth]{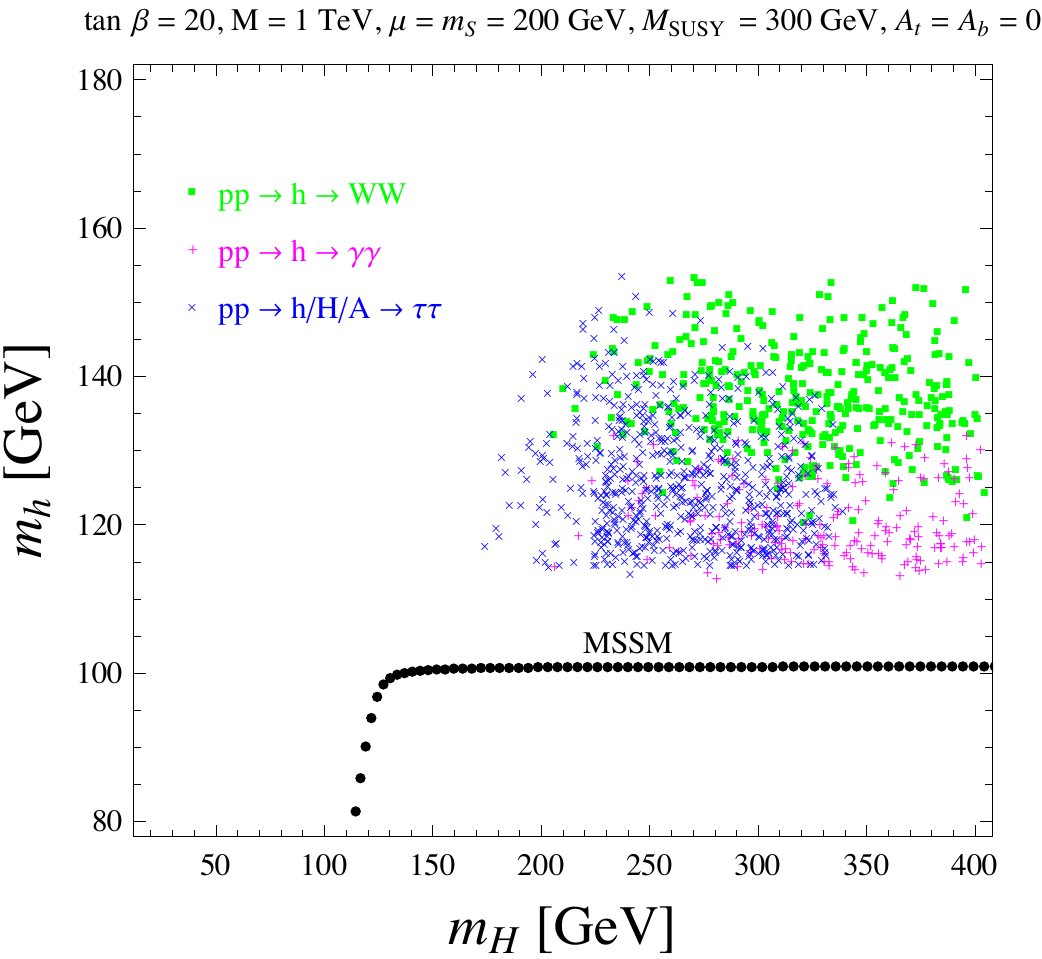}
\newline
\end{center}
\end{minipage}
\caption{\small Scan over BMSSM scenarios, for $\tan\beta=20$.  Upper:
Points not excluded by LEP and Tevatron data at the 95\% CL. We show
points currently excluded by the LHC with 1 fb$^{-1}$ in green.  The
LHC exclusion reach for 5 and 15 fb$^{-1}$ is shown in magenta and
blue, while points outside the LHC-Run I reach are plotted in red
(there are none here).  Lower: Models that can be discovered
(5$\sigma)$ after 15 fb$^{-1}$ of collected data.  We include here
those points that are currently excluded at less than $3\sigma$ (see
text).  We also indicate the discovery mode: $pp \to h \to WW$
(green), $pp \to h \to \gamma \gamma$ (magenta), $pp \to h/H/A \to
\tau^+\tau^-$ (blue).  Note that the color code is different in the
two plots.  }
\label{fig:wwtb2gammagammatb20}
\end{center}
\end{figure}

In Fig.~\ref{fig:wwtb2gammagammatb20} we show the same information as
in Fig.~\ref{fig:mavsmltoday}, but for $\tan \beta=20$.  
One sees in the left panel that the current bounds make sharp cuts on
the parameter space: LEP rules out points with $m_h \lesssim114.4$
GeV, while the Tevatron excludes the heavy mass points, effectively
setting an upper bound on our sample of around 160 GeV. The 7 TeV run
of the LHC can exclude all of our scanned points with a significance
larger than 2$\sigma$ for a total integrated luminosity of about
$16~\rm{fb}^{-1}$.  It is worth stressing that for $\tan \beta=20$
there is a significant number of models being probed by the $\tau^+
\tau^-$ decay mode (about half of the points tested at less than
$3\sigma$ with the ATLAS and CMS analyses of the Summer of 2011).

As for the discovery prospects (right panel of Fig.~\ref{fig:wwtb2gammagammatb20}), we observe that $h \to
WW$ is an important discovery mode for lightest CP-even Higgs masses
heavier than 120 GeV, and for CP-odd masses larger than 200 GeV, and
that the inclusive tau channel is useful for discovery at $m_H < 325$
GeV. For light values of $m_h$ (in the 115-130 GeV range) we have that
the $\gamma \gamma$ channel becomes a discovery mode, playing a more
important role than in the $\tan \beta=2$ case.  This is due to the
enhancement of the Higgs signal in the diphoton channel, which in some
cases can be as large as a factor of 8 above the SM. However, such a
large diphoton signal is excluded by the current LHC dataset, that is
able to test rates between 1.5-3 times the SM after combining the CMS
and ATLAS limits.  We note also that here the $H \to WW$ channel does
not play a role, mainly due to the fact that in the large $\tan \beta$
regime the heavy CP-even Higgs coupling to electroweak vector bosons
tends to be suppressed with respect to the SM value.
We emphasize that essentially all the points in our
scan for $\tan\beta = 20$ can be tested with $15~{\rm fb}^{-1}$.  The
couple of points marked as ``not probed'' in the left panel of
Fig.~\ref{fig:wwtb2gammagammatb20} can actually be probed in the
$\tau^{+}\tau^{-}$ channel at the $2\sigma$ level with the slightly
larger luminosity of~$\sim 16~{\rm fb}^{-1}$.

\begin{figure}[!t]
\begin{center}
\includegraphics[width=0.5\textwidth]{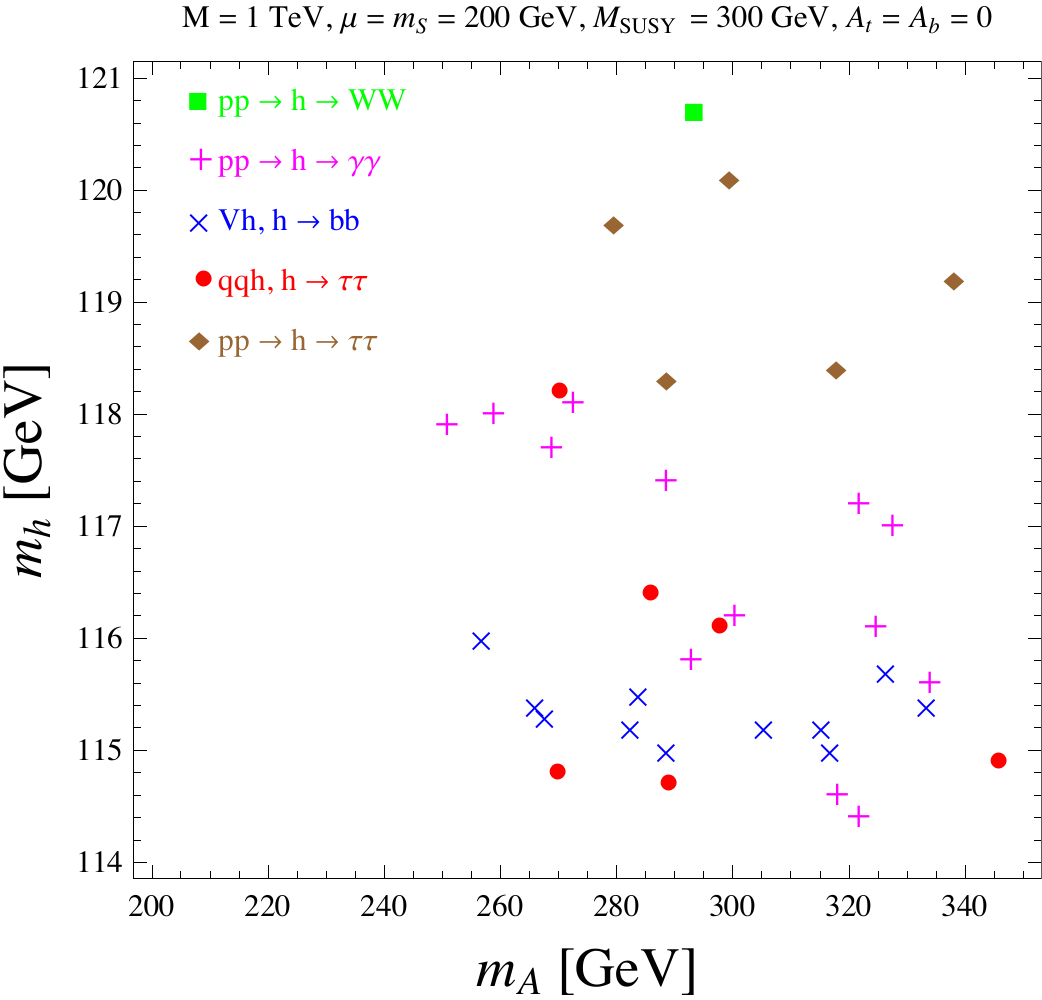}
\caption{Sparse scan over BMSSM scenarios, for $\tan\beta=4,6,8$.  We
show only points that cannot be probed at the $2 \sigma$ level at the
7 TeV LHC after 15 fb$^{-1}$ per experiment of integrated luminosity.
We indicate the discovery mode: $pp \to h \to WW$ (green), $pp \to h
\to \gamma \gamma$ (magenta), $Vh, h \to b \bar{b}$ (blue), $qqh, h
\to \tau^+ \tau^-$ (red) and $pp \to h \to \tau^+\tau^-$ (brown).  }
\label{fig:tbintermediate}
\end{center}
\end{figure}

The LHC has great potential to discover Higgs bosons, but it is also
possible to think of a scenario where, by the end of the 7 TeV LHC
run, the SM Higgs would be excluded in the whole mass range, without
any excess over the expectations in all search channels.  In that
case, almost all of our points for $\tan \beta=2$ and $\tan \beta=20$
would be excluded as well with $15~{\rm fb}^{-1}$.  However, for
intermediate values of $\tan \beta$, where the MSSM searches are less
efficient and the search for $h$ is more challenging (with $m_h$ in
the 114-120 GeV range), one would be left with a fraction of parameter
space not probed with $15~{\rm fb}^{-1}$.  Nevertheless, a sparse scan
over $\tan \beta$ (using values of 4, 6 and 8) suggests that all such
points can be probed with $20~\rm{fb}^{-1}$ of integrated luminosity.
We show these points in Fig.~\ref{fig:tbintermediate}, in the $m_h$--$m_A$
plane, indicating the exclusion channel in each case: $pp \to h \to
WW$ (green), $pp \to H \to \gamma \gamma$ (magenta), $Vh, h \to b
\bar{b}$ (blue), $qqh, h \to \tau^+ \tau^-$ (red) and $pp \to h \to
\tau^+\tau^-$ (brown).

There are channels that have not yet been exploited by the
experimental collaborations, e.g. $h \to AA$, $H \to hh$, or a charged
Higgs search with suppressed $BR(H^+ \to \tau^+ \nu_\tau)$.  Dedicated
searches in these decay modes could provide additional opportunities
for Higgs discovery.

\subsection{Benchmark scenarios}

Considering the Summer 2011 LHC results we update the analysis of the
BMSSM benchmark scenarios presented in Ref.~\cite{Carena:2010cs} and
introduce a number of additional points that illustrate the
possibilities at the LHC with up to 15 fb$^{-1}$ of data.  We keep the
notation of Ref.~\cite{Carena:2010cs} for each point, labeling them
from A to H, and show them as stars marked with the corresponding
letter in Figs.~\ref{fig:mavsmltoday} and
\ref{fig:wwtb2gammagammatb20}.

We recall first our notation and conventions.  The effective couplings
squared, $g^2_{\phi X}$, are computed as the ratio of the partial
widths of the Higgs boson in our model to the SM ones, for $\phi \to
X=gg,WW,...$.  The effective cross section $g^2_{pp \to X}$, on the
other hand, is defined as the ratio of the total \textit{inclusive}
cross section at the LHC in our model normalized to the SM result.
The cross sections in our model are obtained by scaling each
production mode with the corresponding effective coupling squared:
$g^2_{\phi gg}$ for gluon fusion, $g^2_{\phi WW/ZZ}$ for the
Higgs-strahlung and vector boson fusion, and $g^2_{\phi b
\bar{b}}$/$g^2_{\phi t \bar{t}}$ for the associated production with a
bottom/top pair, respectively.  We also define $Q_i ({\cal L})$ as the
ratio between the signal (production cross section times branching
fraction) in our model, in channel $i$, and the LHC 95\% CL limit on
this rate, at the luminosity ${\cal L}$.  For each benchmark point, we
will simply report the maximum value among all the $Q_i$.
We will call by ${\cal L}_{2}$ and ${\cal L}_{5}$ the luminosities
required to claim a 2$\sigma$ exclusion (from now on, exclusion) or a
$5\sigma$ discovery (from now on, discovery).  For further details see
Appendix \ref{app:significances}.

\subsubsection*{Benchmark scenarios for $\tan \beta=2$.}

Of the six low-$\tan\beta$ benchmark points proposed in
Ref.~\cite{Carena:2010cs}, four have been excluded at the $\sim
4\sigma$ level by the most recent LHC studies and two are not yet probed.  The properties of
these points (spectrum, couplings, branching ratios) were presented in
\cite{Carena:2010cs}.  For the excluded ones, here we simply summarize
the exclusion channels and the associated significances.  All these
points are indicated by stars in the $m_{H}-m_{h}$ plane of
Fig.~\ref{fig:mavsmltoday}:\\[0.5em]
\noindent \underline{\textit{Point B}} is excluded independently by
the $pp \to h \to WW$ and $pp \to h \to \gamma\gamma$ LHC searches,
both at the 4$\sigma$ level.  In fact, for this model, which has
$m_{h} \approx 130~{\rm GeV}$, the $pp \to h \to ZZ$ channel also
excludes it at the 2.1$\sigma$ level.  The reason for such a high
exclusion capability from LHC searches is a significant suppression in
the $h \to b\bar{b}$ channel, that results in enhancements of 3.6, 2.7
and 3.5 w.r.t.~SM rates in the $WW$, $\gamma\gamma$ and $ZZ$ channels,
respectively.  \\[0.5em]
\noindent \underline{\textit{Point C}} is excluded by current data in
the WW channel, with a statistical significance of 3.8$\sigma$
(2.4$\sigma$) for $h$ (H), that may be interpreted to give a combined
exclusion at the 4.5$\sigma$ level.  \\[0.5em]
\noindent \underline{\textit{Point D}} is excluded independently by
the $pp \to h \to WW/ZZ$ searches at the 2.5/3.3$\sigma$ level, which
yield a combined exclusion at the 4.1$\sigma$ level.  \\[0.5em]
\noindent \underline{\textit{Point E}} is excluded independently by
the $pp \to h/H \to WW$ searches at the 2.1/1.9$\sigma$ level, and by
the $pp \to H \to ZZ$ search at the 2.7$\sigma$ level, thus yielding a
combined exclusion significance at the 3.9$\sigma$ level.

\medskip \noindent One should notice that these points are strongly
probed by the weak di-boson channels.  This is a rather direct
consequence of the higher-dimension operators, which can have the
following effects:
\begin{itemize}

\item The lightest CP-even Higgs mass can increase sufficiently to
make the $WW$, or even the $ZZ$ decay modes sizable (or dominant).

\item Both CP-even Higgs states can mix significantly so that they can
\textit{both} have sizable couplings to weak gauge boson pairs.
Sometimes, it is the heavier CP-even state that couples dominantly to
$WW$ or $ZZ$.

\end{itemize}
These general observations imply that typically one or the other
CP-even Higgs state (or in some cases both) is constrained by the SM
Higgs searches in the above di-boson channels.  Such a situation is
far less typical in the MSSM, where the lightest CP-even state decays
dominantly into $b\bar{b}$ pairs, and would be searched for more
efficiently in the $\gamma\gamma$ channel (although also in the MSSM,
at low $\tan\beta$, the heavier CP-even $H$ --while having a
suppressed coupling to $VV$-- can still have a sizable decay branching
fraction into weak gauge bosons if its mass is in the appropriate
kinematic range).

\begin{table}[t]
\begin{center}
{\bf POINT $A'$}  
\end{center}
\begin{center}
\begin{tabular}{|c|c|c|c|}
\hline
$m_A ~({\rm GeV})$ & $m_h ~({\rm GeV}) $ & $m_H ~({\rm GeV}) $ & $m_{H^{\pm}} ~({\rm GeV}) $ \\
\hline
129 & 143 & 194 & 148 \\
\hline
$g_{hWW}^2$ & $g_{HWW}^2$  &  $g_{pp \to h}^2$ & $g_{pp \to H}^2$ \\
\hline
0.24   & 0.73 & 1.24  & 0.48 \\
\hline
channel & BMSSM (SM) & channel & BMSSM (SM)  \\ 
\hline
$h \to b \bar{b}$           & 0.62~(0.30)  &$ h \to WW $  & 0.21~(0.55) \\
$H \to WW$                  & 0.74~(0.75)   &
$H \to ZZ$	             & 0.24~(0.25)   \\
$A \to b \bar{b}$	     & 0.89 &
$A \to \tau \bar{\tau}$    & 0.10   \\
$H^{+} \to \bar{\tau} \nu_{\tau}$ & 0.82 & $H^{+} \to t \bar{b}$  & 0.15  \\
\hline
\hline
$p p \to h \to WW $ & Q(15 fb$^{-1}$) & ${\cal L}_2$ (fb$^{-1}$) & ${\cal L}_5$ (fb$^{-1}$) \\
\hline 
0.46 & 2.8 & 1.9 & 11.9 \\
\hline
\end{tabular}
\end{center}
\caption{ {\em Masses and branching fractions in the BMSSM (and in the
SM for $h$ and $H$) for scenario $A'$.  We only show the main decay
modes.  The rate of the most sensitive channel is normalized to the
SM.}}
\label{tab:pointA5}
\end{table}
\begin{table}[t]
\begin{center}
{\bf POINT $B'$} 
\end{center}
\begin{center}
\begin{tabular}{|c|c|c|c|}
\hline
$m_A ~({\rm GeV})$ & $m_h ~({\rm GeV}) $ & $m_H ~({\rm GeV}) $ & $m_{H^{\pm}} ~({\rm GeV}) $ \\
\hline
133 & 117 & 156 & 156 \\
\hline
$g_{hWW}^2$ & $g_{HWW}^2$  &  $g_{pp \to h}^2$ & $g_{pp \to H}^2$ \\
\hline
0.90   & 0.10 & 0.71  & 0.94 \\
\hline
channel & BMSSM (SM) & channel & BMSSM (SM)  \\ 
\hline
$h \to b \bar{b}$           & 0.84~(0.73)  &$ h \to \tau \bar{\tau} $  & 0.09~(0.08) \\
$H \to b \bar{b}$           & 0.64~(0.10)   &
$H \to \tau \bar{\tau}$	             &  0.12~(0.01) \\
$H \to WW$	     & 0.23~(0.80)   &
$A \to b \bar{b}~/~\tau \bar{\tau}$    & 0.89 / 0.10   \\
$H^{+} \to \bar{\tau} \nu_{\tau}$ & 0.72 & $H^{+} \to t \bar{b}$  & 0.24  \\
\hline
\hline
$p p \to H \to WW $ & Q(15 fb$^{-1}$) & ${\cal L}_2$ (fb$^{-1}$) & ${\cal L}_5$ (fb$^{-1}$) \\
\hline 
0.27 & 2.8 & 1.9 & 12.0 \\
\hline
\end{tabular}
\end{center}
\begin{center}
{\bf POINT $C'$}  
\end{center}
\begin{center}
\begin{tabular}{|c|c|c|c|}
\hline
$m_A ~({\rm GeV})$ & $m_h ~({\rm GeV}) $ & $m_H ~({\rm GeV}) $ & $m_{H^{\pm}} ~({\rm GeV}) $ \\
\hline
203 & 118 & 222 & 225 \\
\hline
$g_{hWW}^2$ & $g_{HWW}^2$  &  $g_{pp \to h}^2$ & $g_{pp \to H}^2$ \\
\hline
1.0   & $\le 10^{-3}$  & 1.22  & 0.4 \\
\hline
channel & BMSSM (SM) & channel & BMSSM (SM)  \\ 
\hline
$h \to b \bar{b}$           & 0.70~(0.72)  &$ h \to \tau \bar{\tau} $  & 0.07~(0.07) \\
$h \to WW $                  & 0.13 ~(0.12)  & $ h \to \gamma \gamma ~(\times 10^{-3})$  & 2.1 ~(2.3) \\
$H \to b \bar{b}~/~\tau \bar{\tau} $           & 0.81 / 0.10   &
$H \to WW$	             & 0.04   \\
$A \to b \bar{b}~/~\tau \bar{\tau} $	     & 0.87 / 0.10 &
$H^{+} \to t \bar{b}$      & 1.0  \\
\hline
\hline
$p p \to h \to \gamma\gamma $ & Q(15 fb$^{-1}$) & ${\cal L}_2$ (fb$^{-1}$) & ${\cal L}_5$ (fb$^{-1}$) \\
\hline 
1.1 & 1.8 & 4.8 & 30 \\
\hline
\end{tabular}
\end{center}
\caption{ {\em Masses and branching fractions in the BMSSM for
scenarios $B'$ and $C'$.  We only show the main decay modes.  The rate
of the most sensitive channel is normalized to the SM.}}
\label{tab:pointC5}
\end{table}

The left panel of Fig.~\ref{fig:mavsmltoday} shows that most models
with $m_{h} \gtrsim 135~{\rm GeV}$ are excluded at least at the
$2\sigma$ level by the most recent LHC searches.  Furthermore, most of
the remaining points will be probed with $5~{\rm fb}^{-1}$ of
integrated luminosity (combining both experiments), as shown by the
magenta points in the same plot.  The points not currently excluded
will be tested mainly in the $h\to WW$ channel, but there are also
many models where such a signal would actually correspond to the
heavier CP-even Higgs.  Finally, a few points will be tested in the
$h\to \gamma\gamma$ channel; these have $m_{h} \approx 115-118~{\rm
GeV}$ and $m_{H} > 220~{\rm GeV}$ (with very suppressed
$g^{2}_{HWW}$).  We show in Tables~\ref{tab:pointA5}--\ref{tab:pointF}
representative examples of the models that can be probed by the end of
2011 (i.e., assuming $5~{\rm fb}^{-1}$ of integrated luminosity per
experiment).  

Other benchmark scenarios with neutral Higgs bosons
decaying into diboson and requiring luminosities of
order $10$ fb$^{-1}$ to be probed are present in our scan, and it would be interesting to explore these options in the case of no positive signals by the end of 2011. \\[0.5em]
\noindent \underline{\textit{Point $A'$}} in Table~\ref{tab:pointA5}
corresponds to a model that can be excluded (discovered) in the $pp
\to h \to WW$ channel with $1.9~(11.9)~{\rm fb}^{-1}$.  Note that in
this case, it is actually the heavy CP-even Higgs $H$ that couples
more strongly to the gauge bosons.  With a mass of $m_{H} \approx
194~{\rm GeV}$, the model can be excluded independently at the 95\%~CL
in the $pp\to H\to WW$ and $pp\to H\to ZZ$ channels, with ${\cal L}
\approx 2.6~{\rm fb}^{-1}$ and ${\cal L} \approx 4.5~{\rm fb}^{-1}$,
respectively.  On the other hand, a discovery in these two channels
could be obtained with ${\cal L} \approx 16~{\rm fb}^{-1}$ and ${\cal
L} \approx 28~{\rm fb}^{-1}$, respectively.  \\[0.5em]
\noindent \underline{\textit{Point $B'$}} in the upper part of
Table~\ref{tab:pointC5} corresponds to a model that can be excluded
(discovered) in the $pp \to H \to WW$ channel with $1.9~(12)~{\rm
fb}^{-1}$: although the heavy CP-even Higgs $H$ has a suppressed
coupling to $W$-pairs, the BR$(WW)$ is non-negligible, while the
production is slightly reduced with respect to the SM. The CP-even
Higgs boson $h$ can be probed at the 2$\sigma$ level with ${\cal L}
\approx 22~{\rm fb^{-1}}$ in the $Vh, h \to b \bar{b}$ and in the
$qqh, h \to \tau^+ \tau^-$ channels.  \\[0.5em]
\noindent \underline{\textit{Point $C'$}} in the lower part of
Table~\ref{tab:pointC5} corresponds to a model that can be excluded
(discovered) in the $pp \to h \to \gamma\gamma$ channel with
$4.8~(30)~{\rm fb}^{-1}$, and in the $pp \to h \to WW$ channel with
$6.1~(38)~{\rm fb}^{-1}$ In this example $h$ is essentially SM-like,
although it presents some enhancement in production compared to a SM
Higgs.  The remaining Higgs bosons are likely hard to discover at the
LHC run~I in this low $\tan \beta$ scenario.  For instance, for the
pseudoscalar Higgs A, one would need $23~{\rm fb}^{-1}$ for a
2$\sigma$ exclusion in the $\tau^+ \tau^-$ channel.

\begin{table}
\begin{center}
{\bf POINT $F$} 
\end{center}
\begin{center}
\begin{tabular}{|c|c|c|c|}
\hline
$m_A ~({\rm GeV})$ & $m_h ~({\rm GeV}) $ & $m_H ~({\rm GeV}) $ & $m_{H^{\pm}} ~({\rm GeV}) $ \\
\hline
64 & 135 & 155 & 125 \\
\hline
$g_{hWW}^2$ & $g_{HWW}^2$  &  $g_{pp \to h}^2$ & $g_{pp \to H}^2$ \\
\hline
$\le 10^{-2}$ & 0.99 & 0.59 &  1.14 \\
\hline
channel & BMSSM & channel & BMSSM  \\ 
\hline
$h \to b \bar{b} $     & 0.15  & $ h \to AA$  & 0.84  \\
$ H \to WW$	       & 0.12  & $ H \to AA$  & 0.84  \\
$H \to b \bar{b} $    & 0.02  & $ A \to b \bar{b}~/~\tau \bar{\tau} $ & 0.91 / 0.09  \\
$ H^{+} \to \bar{\tau} \nu_{\tau}$ & 0.56  & $H^{\pm} \to  W^{\pm} + A $ & $0.40$    \\
\hline
\hline
$ p p \to H \to WW$ & Q(15 fb$^{-1}$) & ${\cal L}_2$ (fb$^{-1}$)& ${\cal L}_5$ (fb$^{-1}$) \\
\hline 
0.18 & 1.8 & 4.9 & 30 \\
\hline
\end{tabular}
\end{center}
\caption{ {\em Masses and branching fractions in the BMSSM for
scenario F of Ref.~\cite{Carena:2010cs}.  The rate of the most
sensitive channel is normalized to the SM.}}
\label{tab:pointF}
\end{table}
\medskip \noindent \textit{\underline{Point $F$}} is one of the
benchmark points presented in Ref.~\cite{Carena:2010cs} that have not
been excluded (see Table~\ref{tab:pointF}).\footnote{The other point
at low $\tan\beta$ of Ref.~\cite{Carena:2010cs} that has not been
excluded was labeled $A$ in that reference.  It has MSSM-like
characteristics, with a SM-like Higgs with $m_{h}\approx 118~{\rm
GeV}$ that could be excluded (discovered) in the $\gamma\gamma$
channel with $14~(90)~{\rm fb}^{-1}$.  The non-standard Higgs bosons have
masses of about $240~{\rm GeV}$, and are harder to find at the LHC.}
It has a rather light pseudoscalar Higgs, so that both BR$(h/H \to
AA)$ are sizable.  Therefore, possible search channels could be $b \bar{b}
b \bar{b}$, $b \bar{b} \tau^+ \tau^-$, or $\tau^+ \tau^- \tau^+
\tau^-$.  Aside from these options, which we are not considering here,
the model can be excluded (discovered) in the $pp \to H \to WW$
channel with $4.9~(30)~\rm{fb}^{-1}$.  The charged Higgs search cannot
be applied in a straightforward manner, since $BR(H^+ \to \tau^{+}
\nu_{\tau}) = 0.56$.  A new interesting decay mode for the charged
Higgs opens up in this case: $H^\pm \rightarrow W^\pm A$.

\medskip \noindent Finally, there are a couple of low-$\tan\beta$
models labeled in the left panel of Fig.~\ref{fig:mavsmltoday} as
``not probed'' (red points).  In some case, this is due to the
presence of a relatively light CP-odd Higgs that provides an
additional decay channel for $h$ and/or $H$, which suppresses the
signal in the channels probed so far, thus making them ineffective
even with $15~{\rm fb}^{-1}$.  In other cases, the BR into $b\bar{b}$
presents an enhancement that also has the effect of reducing the
signal in the most sensitive channels.  However, we find that all such
models have a relatively light charged Higgs ($\sim 115-130~{\rm
GeV}$) with a non-negligible branching ratio into $\tau \nu_{\tau}$,
and we expect that the $H^+ \to \tau \nu_{\tau}$ channel should be
effective in discovering such a state.  However, the published
analyses do not apply in a straightforward way since here BR$(H^{+}
\to \tau^{+}\nu_{\tau}) \neq 1$, being diluted by the $H^{+} \to W^{+}
A$ decay channel.

\subsubsection*{Benchmark scenarios for $\tan \beta=20$.}

The two $\tan\beta = 20$ benchmark points that were defined in
Ref.~\cite{Carena:2010cs} have been excluded as follows: \\[0.5em]
\noindent \underline{\textit{Point G}} is excluded by the $pp \to h
\to WW$ searches at the
2.6$\sigma$ level.  \\[0.5em]
\noindent \underline{\textit{Point H}} is excluded by the $pp \to h
\to \gamma\gamma$ searches at the
5.8$\sigma$ level.  \\[0.5em]
We see in the right panel of Fig.~\ref{fig:wwtb2gammagammatb20} that
the most sensitive channels are $pp \to h \to WW$, $pp \to h \to
\gamma\gamma$ and $pp \to h/H/A \to \tau^+\tau^-$ (at large
$\tan\beta$ the $HVV$ couplings are always suppressed).  We select
here additional benchmark points, illustrating the above cases.  These
are models that are presently allowed and can be discovered with
${\cal L} = 15~{\rm fb}^{-1}$: \\[0.5em]
\begin{table}[t]
\begin{center}
{\bf POINT $J$}  
\end{center}
\begin{center}
\begin{tabular}{|c|c|c|c|}
\hline
$m_A ~({\rm GeV})$ & $m_h ~({\rm GeV}) $ & $m_H ~({\rm GeV}) $ & $m_{H^{\pm}} ~({\rm GeV}) $ \\
\hline
302 & 129 & 312 & 305 \\
\hline
$g_{hWW}^2$ & $g_{HWW}^2$  &  $g_{pp \to h}^2$ & $g_{pp \to H}^2$ \\
\hline
0.97   & 0.02 & 1.55  & 0.62 \\
\hline
channel & BMSSM (SM) & channel & BMSSM (SM)  \\ 
\hline
$h \to b \bar{b}$           & 0.65~(0.56)  & $h \to \tau \bar{\tau}$   & 0.11~(0.06) \\
$h \to WW$                   & 0.16~(0.27)  & $H \to b \bar{b}~/~\tau \bar{\tau}$           & 0.65 / 0.11 \\
$H \to h h$	     & 0.13 &   $A \to b \bar{b}~/~\tau \bar{\tau}$    & 0.81 / 0.14   \\
$H^{+} \to \bar{\tau} \nu_{\tau}$ & 0.20 & $H^{+} \to t \bar{b}$  & 0.75  \\
\hline
\hline
$p p \to h \to WW $ & Q(15 fb$^{-1}$) & ${\cal L}_2$ (fb$^{-1}$) & ${\cal L}_5$ (fb$^{-1}$) \\
\hline 
0.93 & 2.7 & 2.1 & 13.1 \\
\hline
\end{tabular}
\end{center}
\caption{ {\em Masses and branching fractions in the BMSSM (and in the
SM for $h$) for scenario $J$.  We only show the main decay modes.  The
rate of the most sensitive channel is normalized to the SM.}}
\label{tab:pointG5}
\end{table}
\noindent \underline{\textit{Point $J$}} in Table~\ref{tab:pointG5}
corresponds to a model that can be excluded (discovered) in the $pp
\to h \to WW$ channel with about $2~(13)~{\rm fb}^{-1}$.  It can also
be excluded (discovered) in $h \to \gamma\gamma$ with $8~(52)~{\rm
fb}^{-1}$.  Note that it has an enhanced branching ratio into $\tau$
pairs (and $b\bar{b}$) compared to the SM, and can be excluded
(discovered) with $4.2~(26)~{\rm fb}^{-1}$ in this channel.  The
non-standard Higgs bosons (around $300~{\rm GeV}$) can be excluded
(discovered) in the $\tau^+\tau^-$ channel with $5.1~(32)~{\rm
fb}^{-1}$ for A and $10.2~(63)~{\rm fb}^{-1}$ for H. Combining their
signals, a total integrated luminosity of $3.4~(21)~{\rm fb}^{-1}$ is
required for exclusion (discovery) .  We also note that since $BR (H
\to hh) \approx 0.13$, one could also look for $H$ by studying the $b\bar{b} \gamma\gamma$, $b
\bar{b} b \bar{b}$, $b \bar{b} \tau^+ \tau^-$, $\tau^+ \tau^- \tau^+
\tau^-$ or even $b \bar{b} W^+ W^-$.
Dedicated studies would be necessary to access the viability of these decay channels
\\[0.5em]
\noindent \underline{\textit{Point $K$}} in Table~\ref{tab:pointH5}
corresponds to a model that can be excluded (discovered) in the $pp
\to h \to \gamma\gamma$ channel with $2.1~(13)~{\rm fb}^{-1}$.  The
non-standard Higgs bosons (around $380~{\rm GeV}$) can be excluded
(discovered) in the $\tau^+\tau^-$ channel with $6.3~(39)~{\rm
fb}^{-1}$ (we add their signals since their mass difference is less
than $10~{\rm GeV}$).  \\[0.5em]
\noindent \underline{\textit{Point $L$}} in Table~\ref{tab:pointL}
illustrate models that can be mainly tested in the $\tau^+\tau^-$
channel.  We note that in such model points only one neutral Higgs
boson can be discovered (in the $\tau^+ \tau^-$ channel) in the 7 TeV
LHC run with about $15~{\rm fb}^{-1}$.  In this example, the pseudoscalar Higgs $A$ can be excluded
(discovered) with $2.2~(13.7)~{\rm fb}^{-1}$, while $h$ and $H$ would
require about $3.5$ and $5.9~{\rm fb}^{-1}$ for a $2\sigma$ excess and
more than 20 fb$^{-1}$ for $5\sigma$, respectively.  In this case, $h$
can also be probed by the $WW$ channel, which requires
$3.6~(22.2)~{\rm fb}^{-1}$ for exclusion (discovery).

\begin{table}[t]
\begin{center}
{\bf POINT $K$}   
\end{center}
\begin{center}
\begin{tabular}{|c|c|c|c|}
\hline
$m_A ~({\rm GeV})$ & $m_h ~({\rm GeV}) $ & $m_H ~({\rm GeV}) $ & $m_{H^{\pm}} ~({\rm GeV}) $ \\
\hline
382 & 115 & 375 & 388 \\
\hline
$g_{hWW}^2$ & $g_{HWW}^2$  &  $g_{pp \to h}^2$ & $g_{pp \to H}^2$ \\
\hline
0.99   & $< 10^{-3}$ & 1.22  & 0.28 \\
\hline
channel & BMSSM (SM) & channel & BMSSM (SM)  \\ 
\hline
$h \to b \bar{b}$           & 0.65~(0.75)  & $h \to \tau \bar{\tau}$   & 0.06~(0.08) \\
$h \to WW$                   & 0.14~(0.08)        & $H \to b \bar{b}~/~\tau \bar{\tau} $           & 0.78 / 0.14 \\
$A \to b \bar{b}$	     & 0.69 & $A \to \tau \bar{\tau}$    & 0.12   \\
$H^{+} \to \bar{\tau} \nu_{\tau}$ & 0.16 & $H^{+} \to t \bar{b}$  & 0.74  \\
\hline
\hline
$p p \to h \to \gamma\gamma$ & Q(15 fb$^{-1}$) & ${\cal L}_2$ (fb$^{-1}$) & ${\cal L}_5$ (fb$^{-1}$) \\
\hline 
1.8 & 2.7 & 2.1 & 13 \\
\hline
\end{tabular}
\end{center}
\caption{ {\em Masses and branching fractions in the BMSSM (and in the
SM for $h$) for scenario $K$.  We only show the main decay modes.  The
rate of the most sensitive channel is normalized to the SM.}}
\label{tab:pointH5}
\end{table}
\begin{table}[t]
\begin{center}
{\bf POINT $L$}  
\end{center}
\begin{center}
\begin{tabular}{|c|c|c|c|}
\hline
$m_A ~({\rm GeV})$ & $m_h ~({\rm GeV}) $ & $m_H ~({\rm GeV}) $ & $m_{H^{\pm}} ~({\rm GeV}) $ \\
\hline
256 & 129 & 278 & 275 \\
\hline
$g_{hWW}^2$ & $g_{HWW}^2$  &  $g_{pp \to h}^2$ & $g_{pp \to H}^2$ \\
\hline
0.99   & 0.02 & 1.6  & 0.86 \\
\hline
channel & BMSSM (SM) & channel & BMSSM (SM)  \\ 
\hline
$h \to b \bar{b}$           & 0.70~(0.56)  & $h \to \tau \bar{\tau}$   & 0.12~(0.06) \\
$h \to WW$         & 0.12~(0.28)         & $H \to  b \bar{b}~/~\tau \bar{\tau}$  & 0.68 / 0.11 \\
$H \to hh $	     & 0.13 &
$A \to b \bar{b}~/~\tau  \bar{\tau}$    & 0.85 / 0.14  \\
$H^{+} \to \bar{\tau} \nu_{\tau}$ & 0.22 & $H^{+} \to t \bar{b}$  & 0.73  \\
\hline
\hline
$\sigma(p p \to X \to \tau \bar{\tau})$ (pb) & Q(15 fb$^{-1}$) & ${\cal L}_2$ (fb$^{-1}$) & ${\cal L}_5$ (fb$^{-1}$) \\
\hline 
$A: $~0.6 & 2.6 & 2.2 & 13.7 \\
$H: $~0.3 & 1.6 & 5.9 & 37 \\
$h: $~2.9 & 2.1 & 3.5 & 22.0 \\
\hline
\end{tabular}
\end{center}
\caption{ {\em Masses and branching fractions in the BMSSM (and in the
SM for $h$) for scenario $L$.  We only show the main
decay modes.  }}
\label{tab:pointL}
\end{table}

Before closing this subsection, we would like to stress the fact that in our $\tan \beta=20$ scan there are points where the lightest CP-even Higgs boson $h$ will be probed with ${\cal O} (10~\rm{fb}^{-1})$ in the $WW$ or $\gamma \gamma$ decay modes, and also points where either $h$, $H$ or $A$ can be tested in the $\tau^+ \tau^-$ channel with a similar luminosity. Those possibilities would be very interesting in the absence of any positive signal with the recently collected ${\cal O} (5~\rm{fb}^{-1})$ data sample.

\subsubsection*{Benchmark scenarios for intermediate $\tan \beta$.}

These scenarios can be probed with about 20 fb$^{-1}$ of integrated
luminosity in the individual $Vh, h \to b \bar{b}$, $qqh, h \to \tau^+ \tau^-$, or $pp \to h \to \gamma\gamma$
channels.  The $pp \to H/A \to \tau^+ \tau^-$ searches are less effective due to the moderate value of $\tan\beta$, and in some cases the non-standard Higgs bosons may remain beyond the LHC reach. However, we point out that when kinematically open, the $H \to hh$ channel can have a sizable branching fraction, thus giving a potential handle on the extended Higgs sector (beyond the SM-like Higgs). By contrast, this can happen in the MSSM only for significantly smaller values of $\tan\beta$, and would point to the presence of heavy physics, as studied here. As an illustration of a ``more challenging'' scenario, we present

\noindent \underline{\textit{Point $M$}} in Table~\ref{tab:pointM}, which 
illustrates intermediate $\tan\beta$ ($\sim 6$) models that require more than $15~{\rm fb}^{-1}$ for exclusion (we have allowed here for $A_{t} = A_{b} = 300~{\rm GeV}$, keeping $M_{\rm SUSY} = 300~{\rm GeV}$).  Note that the diphoton channel is suppressed by a factor of almost 2 compared to the SM, and would require $\sim 20~\rm{fb}^{-1}$ for exclusion. The lightest CP-even Higgs could be excluded earlier in the 
$Vh, h \to b \bar{b}$ or $qqh, h \to \tau \tau$ channels with
$\sim 16~\rm{fb}^{-1}$ and $\sim 17~\rm{fb}^{-1}$, respectively.  However, naively combining the three
search channels one could achieve exclusion with only $\sim 6~{\rm
fb}^{-1}$.  

\begin{table}[t]
\begin{center}
{\bf POINT $M$}   
\end{center}
\begin{center}
\begin{tabular}{|c|c|c|c|}
\hline
$m_A ~({\rm GeV})$ & $m_h ~({\rm GeV}) $ & $m_H ~({\rm GeV}) $ & $m_{H^{\pm}} ~({\rm GeV}) $ \\
\hline
200 & 115 & 193 & 203 \\
\hline
$g_{hWW}^2$ & $g_{HWW}^2$  &  $g_{pp \to h}^2$ & $g_{pp \to H}^2$ \\
\hline
0.99   & $\le 10^{-2}$  & 1.12  & 0.37 \\
\hline
channel & BMSSM (SM) & channel & BMSSM (SM)  \\ 
\hline
$h \to b \bar{b}$  & 0.82~(0.75)  & $h \to \tau \bar{\tau}$   & 0.09~(0.08) \\
$h \to WW$  & 0.05 (0.09) & $H \to  b \bar{b}~/~\tau \bar{\tau}$    &  0.85 / 0.11  \\
$A \to b \bar{b}~/~\tau  \bar{\tau}$    & 0.88 / 0.12 &   $H^{+} \to \bar{\tau} \nu_{\tau} / t \bar{b} $ & 0.25 / 0.74 \\
\hline
$Vh, h \to b \bar{b} $ & Q(15 fb$^{-1}$) & ${\cal L}_2$ (fb$^{-1}$) & ${\cal L}_5$ (fb$^{-1}$) \\
\hline 
1.08 & 0.98 & 15.6 & 98 \\
\hline
$qqh, h \to \tau \bar{\tau} $ & Q(15 fb$^{-1}$) & ${\cal L}_2$ (fb$^{-1}$) & ${\cal L}_5$ (fb$^{-1}$) \\
\hline 
1.12 & 0.94 & 17 & 106 \\
\hline
$pp \to h \to \gamma\gamma $ & Q(15 fb$^{-1}$) & ${\cal L}_2$ (fb$^{-1}$) & ${\cal L}_5$ (fb$^{-1}$) \\
\hline 
0.56 & 0.83 & 22 & 137 \\
\hline
\end{tabular}
\end{center}
\caption{ {\em Masses and branching fractions in the BMSSM (and in the
SM for $h$) for scenario $M$.  We only show the main decay modes.  }}
\label{tab:pointM}
 \end{table}
%
\section{Application to specific UV theories}
\label{sec:UV}

The analysis of the previous sections relies solely on the inclusion
of higher-dimension operators involving the MSSM Higgses ($H_{u}$ and
$H_{d}$) in the super- and K\"ahler potentials, suppressed by up to
$1/M^2$, where $M$ is the scale of the physics being integrated out.
As mentioned in the introduction, our formalism assumes that these are
nearly supersymmetric thresholds, thus treating SUSY breaking in the
heavy sector as a perturbation (included via a spurion superfield).
We emphasized in Ref.~\cite{Carena:2009gx} that the generic EFT
operators can be obtained from extensions of the MSSM involving
massive singlet and triplet superfields, as well as massive gauge
fields ($W'$ and/or $Z'$).  However, there are two classes of
operators that seem hard to induce at tree-level [those with
coefficients proportional to $c_{6}$ and $c_{7}$ in
Eq.~(\ref{eq:DeltaKSUSY})], but that are allowed by supersymmetry.
Although we allowed in our scan that all of these operators have
order-one coefficients, it often happens that turning-off $c_{6}$ and
$c_{7}$ results in changes to the spectrum that are within the
uncertainties expected in the EFT. One can therefore get an idea of
the type of new physics that could be associated with a given EFT
benchmark point by using the relations derived in~\cite{Carena:2009gx}
between such UV examples and the EFT. We have done this explicitly for
the benchmark points of the previous section, and present some of the
details (and caveats) in Appendix~\ref{app:UV}.  This serves as a
``proof of existence'' that the qualitative physics studied within the
EFT can be obtained in specific (even if complicated) UV
completions.\footnote{Such UV completions may in turn have Landau
poles at some intermediate scale, which would indicate the presence of
additional, much heavier physics.  Such issues do not concern us in
this work, since their effects on the Higgs sector can be expected to
be suppressed.}

In this section we illustrate a somewhat orthogonal aspect.  We focus
on specific ``simple'' UV extensions of the MSSM, and perform a study
of the corresponding bounds from the $7~{\rm TeV}$ run of the LHC,
within the EFT framework.  In these cases, we can further analyze the
physics in the full model (e.g.~the spectrum), and in this way
quantify the uncertainties in the EFT. We will focus on two classes of
models: extensions by a massive singlet, and extensions by massive
$SU(2)_{L}$ triplets.  We do not consider gauge extensions, since they
necessitate adding a suitable sector that breaks the extended gauge
symmetry to the SM one, and this sector can give further contributions
to the EFT operators.  Although such an analysis could be done in
principle, the results would be much more model-dependent.  At any
rate, we find that the EFT is in good agreement with the predictions
of the UV theory in the simpler cases we analyze in the following two
subsections, which provides confidence for the generic EFT results.

\subsection{Singlet models}
\label{sec:Singlet}

\begin{figure}[!t]
\begin{center}
\begin{minipage}[b]{0.45\linewidth}
\begin{center}
\includegraphics[width=1\textwidth]{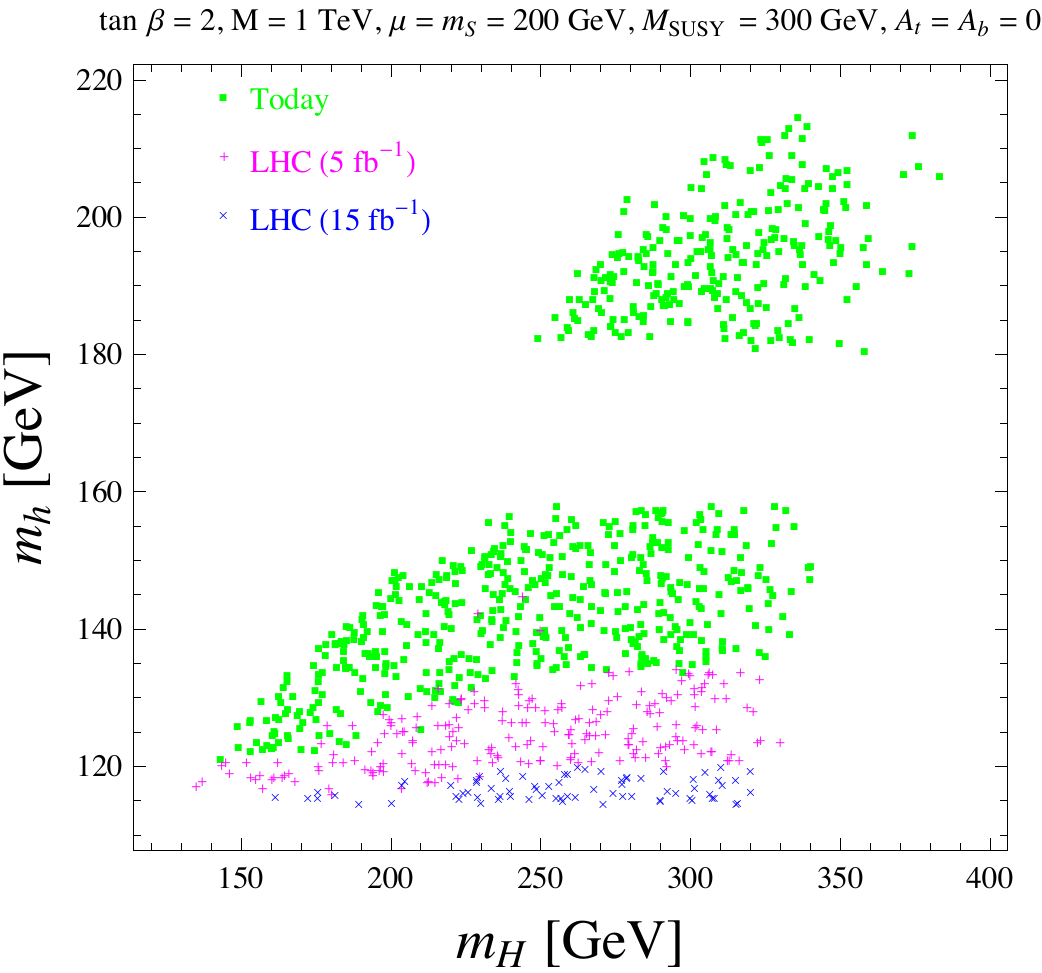}
\newline
(a)
\end{center}
\end{minipage}
\hspace{0.5cm}
\begin{minipage}[b]{0.45\linewidth}
\begin{center}
\includegraphics[width=1\textwidth]{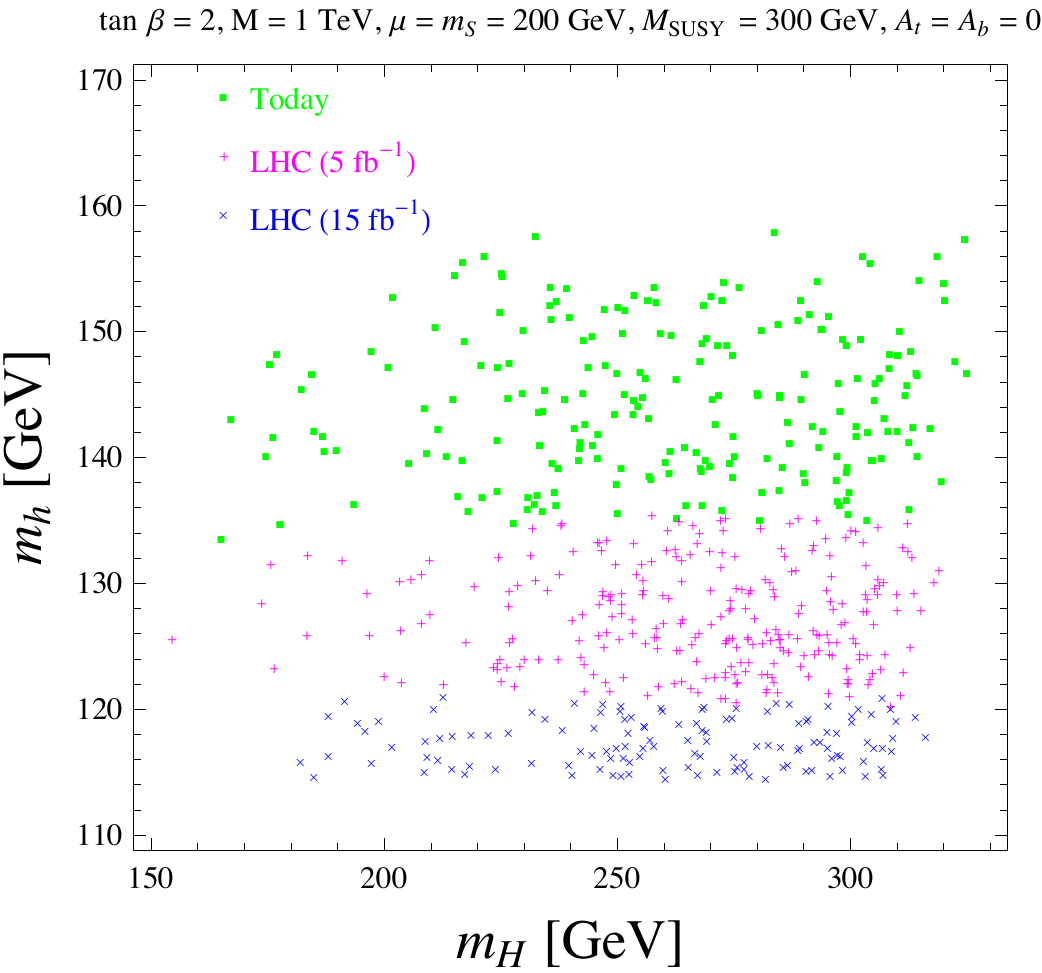}
\newline
(b)
\end{center}
\end{minipage}
\caption{Left panel: scan over parameter points in the singlet theory
of Eq.~(\ref{eq:singlet}), showing the current LHC sensitivity and
projections for the $7~{\rm TeV}$ LHC run.  Right panel: scan
corresponding to a theory extended by $SU(2)_{L}$ triplets, as defined
in Eq.~(\ref{eq:triplets}), showing the current LHC sensitivity and
projections for the $7~{\rm TeV}$ LHC run.  Both examples correspond
to $\tan\beta=2$.}
\label{fig:UVModels}
\end{center}
\end{figure}

Consider a model where the MSSM is extended by a singlet superfield, $S$,
with the following superpotential (apart from the standard Yukawa
interactions):~\footnote{For convenience in the numerical analysis we
have flipped the sign of $M$ compared to Ref.~\cite{Carena:2009gx}.}
\begin{eqnarray} 
W &=& \mu H_u H_d - \frac{1}{2} M S^2 + \lambda_S S H_u H_d  - X \left( \frac{1}{2} a_2 M S^2 + a_3 \lambda_{S} S H_u H_d  \right)~,
\label{eq:singlet}
\end{eqnarray}
where $a_{2}$ and $a_{3}$ are dimensionless, and $X = m_{s} \theta^2$
is a (dimensionless) spurion superfield parameterizing SUSY breaking
in the singlet sector.  We also add the usual non-holomorphic masses
for $H_{u}$, $H_{d}$ and $S$ (the latter taken to be $m_{s}^2$), as
well as the standard $b$-term.  Integrating out $S$ at tree-level
induces the following coefficients in the effective theory (see
Eq.~(\ref{eq:DeltaKSUSY}) and Ref.~\cite{Carena:2009gx} for the
definitions of the coefficients in the EFT, $\omega_{1}$,
$\alpha_{1}$, $c_{i}$, $\gamma_{i}$, $\beta_{i}$):
\be
\begin{split}
& \omega_{1} = \lambda_S ^2 ~, \qquad \hspace{2.8mm}  \alpha_1 = a_{2} - 2 a_{3}~,
\\
& c_4 =  |\lambda_S|^2 ~, \qquad \gamma_4 = a_{2} - a_{3}~, \qquad \beta_4 =  \left| a_{2} -  a_{3} \right|^2  - 1~.
\end{split}
\ee
We have scanned over $\lambda_{S}$, $a_{2}$ and $a_{3}$ (allowing
$\lambda_{S}$ to be as large as 1.5), and performed the EFT checks
described in~\cite{Carena:2009gx} (i.e.~those used in the generic
analysis of Section~\ref{sec:results}).  We have also fixed the SUSY
spectrum (the parameter $M_{\rm SUSY}$ describing the stop sector) as
in the previous section.  The results for $\tan\beta=2$ are shown in
the left panel of Fig.~\ref{fig:UVModels}, showing the power of the
current LHC bounds and the projections for $5$ and $15~{\rm fb}^{-1}$
of integrated luminosity.  We see that the pattern is qualitatively
similar to the one displayed in the model-independent analysis of
Section~\ref{sec:results} (here we scanned up to $m_{A} = 300~{\rm
GeV}$).  In the plot we have explicitly excluded a number of points
with a light CP-odd Higgs (such that either $h \to AA$ or $H \to AA$
are kinematically allowed) for the reasons explained in the next
paragraph.

We have also analyzed exactly the spectrum of states without
integrating out the singlet.  The comparison to the EFT spectrum is
performed by requiring that $v = 174~{\rm GeV}$, $\tan\beta$ and
$m_{H^{\pm}}$ match in the effective and full theories (by adjusting
$m^{2}_{H_{u}}$, $m^{2}_{H_{d}}$ and $b$).  All other parameters
($\mu$, $M$, $\lambda_{S}$, $a_{2}$ $a_{3}$, and $M_{\rm SUSY}$) are
kept fixed.  We choose to match onto the charged Higgs mass (as
opposed to $m_{A}$, for instance) because the charged sector is common
to both theories, and the comparison is therefore cleaner.  We find
that typically the agreement in $m_{h}$ is within 10\% and in $m_{H}$
it is within a few percent.  The largest discrepancies appear in
$m_{A}$ when this state is light, and can reach order 30\%.  One
should then take into account that points where $m_{A}$ is
sufficiently light that the $h \to AA$ channel is open within the EFT,
might get corrections that can change this conclusion.  Except in such
extreme cases (a relatively small number of points), the uncertainties
are as expected in the EFT analysis.  However, we note that in certain
regions a 10\% variation in $m_{h}$ can be relevant
phenomenologically.  Such a change can nevertheless be compensated by
radiative corrections, without affecting too strongly the other
Higgs bosons, and therefore the phenomenological conclusions for the
generic EFT points in our sample can be reasonably obtained within a
singlet extension.  Thus, except possibly for parameter points with
``light'' states, we conclude from this exercise that the analysis
based on the EFT is reliable.  In particular, we trust the results for
the MSSM Higgs sector effective couplings which is much harder to
analyze in the full theory (and is one place where the EFT analysis
shows its power).
%

\subsection{Extensions with Triplets}
\label{sec:Triplets}

Now we consider an extension by $SU(2)_{L}$ triplets.  Specifically,
we include a triplet $\tilde{T}$, with hypercharge $Y=0$, and a
vector-like pair $T$ and $\bar{T}$, with hypercharges $Y = -1$ and $Y
= +1$, respectively.  The superpotential is:~\footnote{Note that, due
to the factor of $1/2$ in $T \equiv T^{a} \tau^{a}$, etc., the
normalizations of the Yukawa couplings are such that one should
``compare'' $\lambda_{\tilde{T}}/2$, $\lambda_{T}/2$ and
$\lambda_{\bar{T}}/2$ to the singlet coupling $\lambda_{S}$ of the
previous subsection.}
\bea
W &=& \mu H_u H_d - \frac{1}{2} M_{\tilde{T}} \tilde{T}^{a} \tilde{T}^{a} + M_T  T^{a} \bar{T}^{a} + \tilde{\lambda}_T H_u \tilde{T} H_d + \frac{1}{2} \lambda_T H_u T H_u + \frac{1}{2} \lambda_{\bar{T}} H_d \bar{T} H_d   \nonumber \\ 
&& \mbox{} - X \left( \frac{1}{2} \tilde{a}_2 M_{\tilde{T}}  \tilde{T}^{a} \tilde{T}^{a} + \tilde{a}_3 \tilde{\lambda}_T H_u \tilde{T} H_d + a_2 M_T  T^{a} \bar{T}^{a} + \frac{1}{2} a_3 \lambda_T H_u T H_u + \frac{1}{2} a_4 \lambda_{\bar{T}} H_d \bar{T} H_d \right)~,
\label{eq:triplets}
\eea
together with non-holomorphic masses for $H_{u}$, $H_{d}$,
$\tilde{T}$, $T$ and $\bar{T}$.  In our scan we will take the soft
masses for all the triplets to be given by $m^{2}_{s}$.  The reason we
include the various triplets simultaneously is that the $Y=0$ triplet
contributes to the Peskin-Takeuchi $T$-parameter with opposite sign to
the vector-like pair with $Y=\pm 1$ (the former contribution is
positive while the latter are negative).  As a result there can
naturally exist partial cancellations in the $T$-parameter that can
allow the parameters to be larger and affect the MSSM Higgs sector
more significantly.

To second order in $1/M_{\tilde{T}} \sim 1/M_{T}$ the contributions to
the EFT operators generated by integrating out the triplets are simply
additive, and were given in~\cite{Carena:2009gx}.  Assuming, for
concreteness, that $M_T = M_{\tilde{T}}$, and following the notation
introduced in Ref.~\cite{Carena:2009gx} and summarized around
Eq.~(\ref{eq:DeltaKSUSY}) , the following coefficients in the EFT are
induced:
\be
\begin{split}
& M = M_T  = M_{\tilde{T}} ~, \qquad \qquad \hspace{-8mm} \omega_{1} = \frac{1}{4} \tilde{\lambda}_T^2 + \frac{1}{4} \lambda_T \lambda_{\bar{T}} ~, \qquad \hspace{4mm}  \alpha_1 = \tilde{a}_{2} - 2 \tilde{a}_{3} + a_{2} - a_{3} - a_{4}~,
\\
& c_1 =  \frac{1}{4} \lvert \lambda_{\bar{T}} \rvert ^2~, \qquad\qquad \gamma_1= a_2-a_4~, \qquad\qquad\qquad \beta_1=  |a_2-a_4|^2 -1~,
\\
& c_2 =  \frac{1}{4} \lvert \lambda_T \rvert ^2~, \qquad\qquad \gamma_2= a_2-a_3~, \qquad\qquad\qquad \beta_2= |a_2-a_3|^2 - 1~,
\\
& c_3 = \frac{1}{2} |\tilde{\lambda}_T|^2 ~, \qquad\qquad \gamma_3 = \tilde{a}_{2} - \tilde{a}_{3}~, \qquad\qquad\qquad \beta_3 =  \left| \tilde{a}_{2} -  \tilde{a}_{3} \right|^2  - 1~,
\\
& c_4 = -\frac{1}{4} |\tilde{\lambda}_T|^2 ~, \qquad\qquad  \hspace{-3mm} \gamma_4 = \tilde{a}_{2} - \tilde{a}_{3}~, \qquad\qquad\qquad \beta_4 =  \left| \tilde{a}_{2} -  \tilde{a}_{3} \right|^2  - 1~.
\end{split}
\ee
We have scanned over $\lambda_{\tilde{T}}$, $\lambda_{T}$,
$\lambda_{\bar{T}}$, $\tilde{a}_{2}$, $\tilde{a}_{3}$, $a_{2}$,
$a_{3}$ and $a_{4}$ (again allowing the $\lambda_{i}$'s to be as large
as 1.5).  However, we keep only points such that $\omega_{1} < 2$ and
$\alpha_{1} < 1.5$ so as to remain within the perturbative regime in
the EFT. We performed again the EFT checks described
in~\cite{Carena:2009gx}, which include checking consistency with EWPT
(allowing for a potential contribution to the oblique parameters from
the SUSY sector, e.g. from splittings in the slepton doublets).  We
have also fixed the SUSY spectrum as in the previous section (and as
in the singlet model above).  The results for $\tan\beta=2$ are shown
in the right panel of Fig.~\ref{fig:UVModels}, showing the power of
the current LHC bounds and the projections for $5$ and $15~{\rm
fb}^{-1}$ of integrated luminosity.  Again, we see that the pattern is
qualitatively similar to the one displayed in the model-independent
analysis of Section~\ref{sec:results}, except that $m_{h}$ reaches
only values of order $160~{\rm GeV}$ (for the range of parameters
described above).  The scan here corresponds to $70~{\rm
GeV}<m_{A}<300~{\rm GeV}$.

As in the case of the singlet model, we have compared the EFT
predictions to the exact spectrum for Eq.~(\ref{eq:triplets}).  We
match again to $v$, $\tan\beta$ and $m_{H^{\pm}}$ by adjusting
$m^{2}_{H_{u}}$, $m^{2}_{H_{d}}$ and $b$.  All other parameters
($\mu$, $M$, $\lambda_{\tilde{T}}$, $\lambda_{T}$,
$\lambda_{\bar{T}}$, $\tilde{a}_{2}$, $\tilde{a}_{3}$, $a_{2}$,
$a_{3}$, $a_{4}$, and $M_{\rm SUSY}$) are kept fixed.  We find that
for the bulk of the scanned points the agreement is within
$10\%$,\footnote{In some cases, the EFT can overestimate $m_{h}$ by as
much as 10\%, which can easily be compensated by radiative
corrections, so that the phenomenological conclusions remain valid.}
and often much better.  Therefore, as for the singlet theory, we
conclude that the EFT analysis captures the physics of the triplet
model reliably.

\section{Conclusions}
\label{sec:conclu}

In this work we have analyzed the current LHC constraints on a large
class of extended Higgs sectors in supersymmetric theories, where the
physics beyond the MSSM (assumed to be approximately supersymmetric)
is somewhat heavier than the MSSM Higgs degrees of freedom.  In order
to perform a relatively model-independent study, we have parameterized
the effects of the extended sector on the MSSM Higgs bosons via
higher-dimension operators.  We consider operators up to dimension-six
in the superpotential and K\"ahler potential, which were shown in
Ref.~\cite{Carena:2009gx} to be potentially very relevant in
determining the phenomenology of the Higgs sector.  In particular, it
was shown in~\cite{Carena:2010cs} that the SUSY Higgs signals could be
markedly different from the standard expectations built on the MSSM
intuition.  The profound distortion of the two-Higgs doublet sector
could have led to striking signals during the very early LHC era.
However, as shown in this work, such scenarios are now highly
constrained by the current null results from the LHC Higgs searches, based on about
1-2~fb$^{-1}$ of integrated luminosity.

This does not mean that extended SUSY Higgs sectors are close to being ruled out, but
it suggests that the Higgs phenomenology will likely be similar to the
MSSM one, with a light CP-even Higgs decaying dominantly into bottom
pairs, that can be searched for more effectively in the diphoton or $W^+W^-$ channel (depending on its mass), and perhaps in the $\tau^{+} \tau^{-}$ channel.  Of course,
non-standard decays into new light states, not considered in this work,
remain also as a possibility.  The non-standard Higgs bosons will likely
have suppressed couplings to the weak gauge bosons, and be somewhat
harder to discover unless $\tan\beta$ is relatively large.  Nevertheless, here can
still exist interesting decays such as $H \to hh$ (that are typically
suppressed in the MSSM, unless $\tan\beta$ is small and in some
tension with the Higgs LEP bound), that can occur with a sizable
branching fraction as a result of the presence of the heavy physics.
Such a signal could indicate the presence of BMSSM physics that could be connected to additional contributions
to the mass of the lightest CP-even, SM-like Higgs state, that may
alleviate the tensions present in the MSSM, as also discussed recently
within specific UV extensions of the MSSM
in~\cite{Delgado:2010uj} and~\cite{Agashe:2011ia}.  We have also analyzed here,
within the effective theory formalism, specific theories involving
additional singlets or $SU(2)_{L}$ triplets, and verified that the EFT
analysis can indeed provide a reasonable approximation in such cases.
It is found that the qualitative conclusions are similar to those
obtained without the prejudice of specific UV theories.

Finally, we have presented projections for 5 and 15~fb$^{-1}$ of integrated
luminosity (roughly anticipating the situation by the end of 2011, and
by the end of the 7 TeV Run of the LHC, respectively), taking into
account the possible combination of ATLAS and CMS results in each of the
relevant search channels.  We find
that at both low (i.e.~order one) and large $\tan\beta$, such amount
of data can either exclude a very large region of parameter space, or
make a discovery.  We also point out that if the SM Higgs is excluded
over the whole mass range, this may be an indication of a Higgs with a
mass close to the LEP bound and an enhancement in the $b\bar{b}$
channel, as can happen both in the MSSM and in some extensions of the
type studied in this work.  Nevertheless, we expect that if
supersymmetry is relevant at the weak scale, signals from its Higgs
sector are likely to appear over the next few years.

\section*{Acknowledgments}
We thank Patrick Draper and Carlos Wagner for useful discussions.
Fermilab is operated by Fermi Research Alliance, LLC under Contract
No.  DE-AC02- 07CH11359 with the U.S. Department of Energy.  EP is
supported by the DOE grant DE-FG02-92ER40699.  J.Z is supported by the
Swiss National Science Foundation (SNF) under contract 200020-138206.
This work was supported by the Research Executive Agency (REA) of the
European Union under the Grant Agreement number PITN-GA-2010-264564
(LHCPhenoNet).  M.C.~would like to thank the Aspen Center for Physics,
where part of this work has been done.

\begin{appendix}

\section{Benchmark Points and UV completions}
\label{app:UV}

In this appendix we establish a connection between the generic scan
over parameters in the EFT, and possible UV completions that could
give rise to such effects.  We illustrate the point with the benchmark
models of Section~\ref{sec:results}.  As mentioned in
Section~\ref{sec:UV}, we can reproduce most of the EFT coefficients by
a combination of massive singlets and triplets, as in the models of
Subsections~\ref{sec:Singlet} and \ref{sec:Triplets}, plus a massive
$W'$ with gauge coupling $\tilde{g}$ and a massive $Z'$ with gauge
coupling $g'$.  For additional details on the EFT operators induced by
these gauge extensions, we refer the reader to our previous
work~\cite{Carena:2009gx}.  In such gauge extensions, there can exist
additional contributions to the EFT operators from the sector that
breaks the extended gauge symmetry to the SM one, if that sector
interacts with the MSSM Higgs fields in any relevant way.  To be
definite, we will assume that any such couplings are subdominant,
since our point here is to \textit{illustrate} that specific generic
benchmark models can actually be obtained from a well-defined theory.
Note that additional contributions give more freedom to obtain a given
set of EFT coefficients, so allowing such couplings would only
strengthen our point.  However, there are a number of issues that
should be taken into consideration:
\begin{itemize}

\item The coefficients proportional to $c_{6}$ and $c_{7}$ cannot be
easily obtained at tree-level, although we have allowed them with
order-one strength in our scan (they are certainly allowed by the
symmetries).  Nevertheless, often their effects do not change the
qualitative features, e.g.~they do not induce changes larger than the
uncertainties already expected in the EFT approximation (especially at
low $\tan\beta$; at large $\tan\beta$ they can be more important).
Also, sometimes those changes can be partially compensated by other,
unrelated effects such as somewhat different radiative corrections.
Thus, it is still interesting to specify what kind of physics could
generate the operators other than those associated with $c_{6}$ and
$c_{7}$.

\item Besides a sector that breaks the extended gauge symmetry to the
SM, in the case of $U(1)'$ additional matter may be necessary to
cancel anomalies.  We do not address this issue here, but note that
the additional matter (probably with masses of order $M$) can give
additional contributions to the EFT operators, again allowing
additional freedom in generating the given EFT coefficients.  However,
for illustration purposes, we will assume that the possible couplings
of these fields to the MSSM Higgs bosons are small.

\item In principle, the different ``heavy'' fields can have somewhat
different masses.  For concreteness, here we will assume a common
(SUSY) mass $M$ for all the heavy states.

\end{itemize}

In Table~\ref{table:UVtanb} we give examples of values of parameters
in UV completions with the above ingredients (singlets, triplets and
gauge extensions) that reproduce all coefficients of the benchmark
scenarios shown in Section~\ref{sec:results} that have not been
excluded by current collider data, except for $c_{6}$ and $c_{7}$
(whose values we also list).  There are more UV parameters than those
in the EFT and therefore some amount of redundancy is present.  We
have arbitrarily fixed the $U(1)'$ charges of $H_{u}$ and $H_{d}$
(denoted by $Q_{u}$ and $Q_{d}$ in the tables), as well as the $U(1)'$
gauge coupling $g'$.  We do not exhibit the parameters associated with
SUSY breaking operators, but there is more than enough freedom in the
UV theory to accommodate those.  It turns out that in most of these
benchmark points one could turn off $c_{6}$ and $c_{7}$ without
changing the conclusions.  An exception is illustrated by Point~K,
where these operators give a positive contribution to $m_{h}$ of about
35\%.  Since Point~K had $m_{h} \sim 115~{\rm GeV}$, turning off
$c_{6}$ and $c_{7}$ would make such a point excluded by LEP. However,
larger radiative corrections than we have assumed could be present,
thus compensating the contribution from $c_{6}$ and $c_{7}$.  A
similar issue is present in most of the $\tan\beta = 20$ examples, but
not in the $\tan\beta = 2$ ones.

We also note that we have assumed that $M = 1~{\rm TeV}$, while the
current bounds on a $W'$ with sequential SM couplings are about
$2.15~{\rm TeV}$ from ATLAS~\cite{Aad:2011yg} and $2.27~{\rm TeV}$
from CMS~\cite{CMS:Wprime}.  Similarly, the current bound on a $Z'$
with SM couplings is $1.83~{\rm TeV}$ from
ATLAS~\cite{Collaboration:2011dca} and $1.94~{\rm TeV}$ from
CMS~\cite{CMS:Zprime}.  However, in our examples the new $SU(2)$
coupling $\tilde{g}$ is fairly suppressed, and in some cases vanishes.
Similarly, the bounds on $Z'$ depend on its couplings to the first two
generations of quarks, which are not constrained by our analysis.
Therefore, we conclude that such UV completions, with gauge resonances
at a TeV, are not necessarily inconsistent with present direct bounds.
\begin{table}[t]
\begin{center}
\begin{tabular}{|c|c|c|c|c|c|c|c|c|c||c|c|}
\hline
\rule{0mm}{5mm}
Point & $\tan\beta$ & $\lambda_{S}^2$ & $\tilde{\lambda}_{T}^2$ & $\lambda_{T}$ & $\lambda_{\bar{T}}$ & $\tilde{g}^2$ & $g^{\prime 2}$ & $Q_{u}$ & $Q_{d}$ & $c_{6}$ & $c_{7}$
\\[0.4em]
\hline
\rule{0mm}{5mm}
$F$ & 2 & 0.17  & -1.7 & 1.8  & 2.8  & 0 & 1.1 & 1/2 & 1/2 & 0.89 & -0.08 
\\[0.4em]
\hline
\rule{0mm}{5mm}
$A'$ & 2 & 0.36 & 0.01 & 1.3 & 1.5 & 0.28 & 0.49 & 1/2 & 1/2 &  -0.38 & -0.29
\\[0.4em]
\hline
\rule{0mm}{5mm}
$B'$ & 2 & -0.24 & 0.37  &  1.2 & 1.4 & 0 & 0.86 & 1/2  & 1/2  & -0.31 & 0.47
\\[0.4em]
\hline
\rule{0mm}{5mm}
$C'$ & 2 &  0.21 & 0.52 & 0.72 & 1.0 & 0 & 0.08 & 1/2 & 1/2 & -0.34 & 0.42
\\[0.4em]
\hline
\hline
\rule{0mm}{5mm}
$J$ & 20 & 0  & -2 &  2.5 & 1.6 & 0  & 1.6  & 1/2  & 1/2  & -0.35 & -0.18
\\[0.4em]
\hline
\rule{0mm}{5mm}
$K$ & 20 & -0.34 & 2.1  & -1.9 & 0.13 & 0.25 & 0.66 & 1/2  & 1/2  & -0.59 & 0.66
\\[0.4em]
\hline
\rule{0mm}{5mm}
$L$ & 20 & 0.29& 2.0 & 1.8 & 0.4 & 0 & 0.88 & 1/2 & 1/2 & 0.61 & -0.92
\\[0.4em]
\hline
\hline
\rule{0mm}{5mm}
$M$ & 6 & -0.36  & -0.21  & 1.8  & 0.62 & 0  & 0.87  & 1/2  & 1/2  & 0 & 0 
\\[0.4em]
\hline
\end{tabular}
\end{center}
\caption{Examples of UV completions that could lead to the set of
effective operators of our benchmark points.  The
parameters $\lambda_{S}$, $\tilde{\lambda}_{T}$, $\lambda_{T}$ and
$\lambda_{\bar{T}}$ were defined in Eqs.~(\ref{eq:singlet})
and~(\ref{eq:triplets}), and may be complex.  $\tilde{g}$ and $g'$ are the gauge couplings
for a heavy $W'$ and $Z'$, respectively (see
Ref.~\cite{Carena:2009gx}).}
\label{table:UVtanb}
\end{table}%
%

\section{Computing significances}
\label{app:significances}

In this Appendix we record the procedure used in the main text to
compute the required luminosities for exclusion and discovery.  A clear
summary of the statistical details and relevant approximations can be
found in Appendix A of Ref.~\cite{Draper:2009au} (see also
Refs.~\cite{Aad:2009wy,Ball:2007zza}).

As a first step, we compute, for each relevant channel/bin $i$, the
following quantity
\bea
\label{eq:defQ}
Q_i ({\cal L}_0) &=& \frac{R_i^{\rm{mod}}}{R_i^{\rm{exp}} ({\cal L}_{0})}~,
\eea
where $R_i^{\rm{mod}}$ is the rate (i.e.~production cross section
times branching fraction) for channel/bin $i$ in a given model, and
$R_i^{\rm{exp}} $ is the exclusion limit at the 95\% C.L.~on this
rate, as reported by the experimental collaborations with a total
integrated luminosity ${\cal L}_{0}$.  Sometimes the experimental
limit is presented normalized to the SM, or to some other reference
model, in which case $R_i^{\rm{mod}}$ should be normalized in the same
way.

While $R_i^{\rm{mod}}$ does not change with luminosity,
$R_i^{\rm{exp}}$ does.  Under the hypothesis that the model in
question is actually realized by Nature, and in the Gaussian limit,
the significance of a (downward) fluctuation by $S_{i}$, is given by
$n_i=S_i/\sqrt{S_i+B_i}$ (neglecting systematic effects).  Thus, in
this limit, and assuming that the data reflects expected background
only, we simply have
\bea
R_i^{\rm{exp}} ({\cal L}) &\approx&  \frac{R_i^{\rm{exp}} ({\cal L}_{0})}{\sqrt{ {\cal L} / { \cal L   }_0   }}~,
\eea
since both signal and background scale linearly with the total
integrated luminosity.  The exclusion is more stringent if a larger
dataset is used, as expected.

Defining $R_i^{\rm{exp}} ({\cal L}_0) = R_{i,0}$ and $Q_{i,0} =
R_i^{\rm{mod}} / R_{i,0}$ $[=Q_i ({\cal L}_0)]$, one then has that
\bea
Q_{i}({\cal L}) &=& \frac{R_i^{\rm{mod}}}{R_{i,0}} \sqrt{ {\cal L }/{\cal L}_0 } ~=~ Q_{i,0} \sqrt{ {\cal L} /{\cal L}_0 }~.
\eea
Since $Q_{i}({\cal L}) = 1$ corresponds to exclusion at the 95\%
C.L.~in channel/bin $i$, the future projection based on the above
simple scaling indicates that the luminosity required to claim
exclusion (of the given model) at the 95\% C.L.~is given by
\bea
\label{eq:exclusion}
{\cal L}_2 &=& \Bigl( \frac{R_{i,0}}{R_i^{\rm{mod}}} \Bigr)^2 {\cal L}_0 ~=~ \Bigl( \frac{1}{Q_{i,0}} \Bigr)^2 {\cal L}_0~.
\eea

One is also interested in estimating the discovery potential.  Here
one imagines that the (future) data shows a $5\sigma$ excess compared
to the background-only expectation.  In this case, the statistical
significance under the background-only hypothesis, in the limit of a
large number of events, is given by $n_i=S_i / \sqrt{B_i}$ (again
neglecting systematics).  If we use the current expectation for the
background, scaled by $\sqrt{\cal L}$ to estimate the expected
background with the higher luminosity, and assume also that $B_{i} \gg
S_{i}$, we can relate the discovery potential to the quantities for
exclusion defined above, since the measures for exclusion and
discovery significance coincide in this limit.  Thus, if the given
model was indeed realized by Nature, and the future data reflected the
expected rate, one would be able to claim a discovery for a luminosity
given by
\bea
\label{eq:discovery}
{\cal L}_5 &=&  \Bigl(\frac{5}{2} \Bigr)^2  \Bigl( \frac{1}{Q_{i,0}}\Bigr)^2~{\cal L}_{0} ~=~ \frac{25}{4} \, {\cal L}_2~.
\eea
Here we used that, under the above hypothesis, the current exclusion
(based on data that reflect background-only) would correspond to a
$2\sigma$ downward fluctuation, and has $Q_{i,0} = 1$.  In the absence
of such a fluctuation, one would have had a ``$2\sigma$ hint'' with
current data.

Throughout this work we make use of Eqs.~(\ref{eq:exclusion}) and
(\ref{eq:discovery}) to compute the required luminosity for an
exclusion or discovery, respectively.

\end{appendix}


\end{document}